\documentclass[12pt]{article}
\usepackage{subcaption}
\usepackage{changepage}
\usepackage[dvipsnames]{xcolor}
\usepackage{xr}
\usepackage[english]{babel}
\usepackage{amsfonts}
\usepackage{amssymb}
\usepackage{bm}
\usepackage{amsmath}
\usepackage{tikz}
\usetikzlibrary{calc,intersections,shapes,decorations,arrows,arrows.meta,fit,positioning, decorations.pathreplacing}
\usepackage{graphicx}
\usepackage{caption}
\usepackage{adjustbox}
\usepackage{threeparttable}
\usepackage{scrextend}
\usepackage{relsize}
\usepackage{booktabs}	
\usepackage[normalem]{ulem}
\usepackage{enumerate}
\usepackage{bbm}
\usepackage{mathtools}
\usepackage{setspace}
\usepackage{soul}
\usepackage{xspace}
\usepackage{csquotes}

\usepackage{placeins}
\usepackage[footfont=normalsize,font=normalsize]{floatrow}
\usepackage{longtable}
\usepackage{mathabx}
 \usepackage{accents}
\usepackage{float}
\usepackage{fullwidth}
\usepackage{verbatim}
\usepackage{autobreak}
\usepackage[bottom]{footmisc} 
\usepackage[titletoc,title, toc,page]{appendix}
\usepackage{pdflscape}

\usepackage[left=2cm,right=2cm,top=2cm,bottom=2cm]{geometry}
    
	\usepackage[longnamesfirst]{natbib}
	\usepackage{bibunits}
	\defaultbibliography{Bibliography.bib}  
	\defaultbibliographystyle{aer}

\definecolor{cite_blue}{rgb}{0.0, 0.1, 0.3}
\definecolor{BrickRed}{rgb}{0.8, 0.25, 0.33}
	\usepackage{hyperref}
	\hypersetup{
    colorlinks=true,
    linkcolor=cite_blue,
    filecolor=cite_blue,      
    urlcolor=cite_blue,
    citecolor = cite_blue
}
	\usepackage{cleveref}

    \usepackage{rotating}
	\usepackage[framemethod=tikz]{mdframed}

    \usepackage{tcolorbox}
    \newtcbox{\feedback}{nobeforeafter,colframe=black,colback=white,boxrule=0.5pt,arc=2pt,
      boxsep=0pt,left=2pt,right=2pt,top=2pt,bottom=2pt,tcbox raise base}
      
    \usepackage{amsthm}
    
    \newtheorem{theorem}{Theorem}[section]
    \newtheorem{assumption}{Assumption}
    
    \newtheorem{prop}{Proposition}[section]
    \newtheorem{lemma}{Lemma}[section]
    
    \theoremstyle{definition}
    \newtheorem{remark}{Remark}[section]
    
    \newtheorem{example}{Example}

\usepackage{ragged2e}
\newlength\ubwidth

\allowdisplaybreaks

\AtBeginDocument{
\setlength{\abovedisplayskip}{6pt}
\setlength{\belowdisplayskip}{6pt}
\setlength{\abovedisplayshortskip}{6pt}
\setlength{\belowdisplayshortskip}{6pt}
}

\deffootnote{1em}{1.6em}{\thefootnotemark \enskip}

\numberwithin{equation}{section}

\renewcommand{\P}{\mathop{}\!\mathbb{P}}

\newcommand{\E}{\mathop{}\!\mathbb{E}}

\newcommand{\N}{\mathcal{N}}

\newcommand{\bracks}[1]{\left[#1\right]}

\newcommand{\expen}[1]{\mathbb{E}_{n}\bracks{#1}}

\newcommand{\parens}[1]{\left(#1\right)}

\newcommand{\convd}{\stackrel{d}{\rightarrow}}

\newcommand{\idx}[1]{\mathbbm{1}\{ #1 \}}

\newcommand{\supp}[1]{\mathrm{Supp}\parens{#1}}

\makeatletter
\newcommand*{\indep}{%
  \mathbin{%
    \mathpalette{\@indep}{}%
  }%
}
\newcommand*{\nindep}{%
  \mathbin{
    \mathpalette{\@indep}{/}%
  }%
}
\newcommand*{\@indep}[2]{%
  \sbox0{$#1\perp\m@th$}
  \sbox2{$#1=$}
  \sbox4{$#1\vcenter{}$}
  \rlap{\copy0}
  \dimen@=\dimexpr\ht2-\ht4-.2pt\relax
  \kern\dimen@
  \ifx\\#2\\%
  \else
    \hbox to \wd2{\hss$#1#2\m@th$\hss}%
    \kern-\wd2 %
  \fi
  \kern\dimen@
  \copy0 
}
\makeatother

\makeatletter
\newcommand*{\addFileDependency}[1]{
  \typeout{(#1)}
  \@addtofilelist{#1}
  \IfFileExists{#1}{}{\typeout{No file #1.}}
}
\makeatother

\begin{document}
\begin{bibunit}

\title{Doubly Robust Estimators with Weak Overlap}

\author{Yukun Ma\thanks{University of Rochester. Email: yma69@ur.rochester.edu} \and Pedro H. C. Sant'Anna\thanks{Emory University. Email: pedro.santanna@emory.edu} \and Yuya Sasaki\thanks{Vanderbilt University. Email: yuya.sasaki@vanderbilt.edu} \and Takuya Ura\thanks{University of California, Davis. Email: takura@ucdavis.edu}}
\date{\today}

\maketitle

\begin{abstract}
Doubly robust (DR) estimators guard against model misspecification but remain sensitive to weak covariate overlap. We show that trimming propensity scores reduces variance but eliminates double robustness. We introduce DR estimators that retain double robustness after trimming through bias correction, preserving the original causal targets across unconfoundedness, instrumental variables, and difference-in-differences designs. In four applications, the proposed estimator yields more precise estimates: ruling out large mortality effects of Medicaid expansion, detecting workforce growth from mental health reform, recovering the Black--White test score gap without strong functional form restrictions, and recovering a positive 401(k) savings effect consistent with the prior literature.

\vspace{4ex}
\end{abstract}

\allowdisplaybreaks
\newpage

\section{Introduction}\label{sec:introduction}

Accounting for confounders is essential for reliable observational causal inference, but doing so poses practical challenges due to model misspecification and weak covariate overlap. Doubly robust (DR) estimators address the first concern: they remain consistent for a prespecified causal estimand even when one component of the working model is incorrect \citep[e.g.,][]{Robins1994}.\footnote{See \cite{Sloczynski2018} for general identification results with doubly robustness in the causal inference.} Yet DR estimators remain vulnerable to the second problem: when propensity scores are close to zero or one, they can become unreliable. A handful of observations with extreme inverse probability weights can dominate the entire estimate, inflating standard errors by factors of two to six, or worse, flipping the sign of point estimates. Weak overlap does not merely reduce precision; it can render observational studies uninformative or misleading.

As we highlight in different applications, this concern is indeed warranted. For instance, we find that the standard DR 95\% simultaneous confidence bands in a staggered difference-in-differences (DiD) analysis of Medicaid expansion on mortality span on average 68 deaths per 100,000 people over the post-treatment horizons---too wide to inform policy. In a 401(k) savings application, the standard DR local average treatment effect (LATE) estimate implies that participation \emph{reduces} net financial assets by \$13,042---a conclusion at odds with the prior literature---because two ineligible observations with extreme instrument propensity scores carry effective weights of roughly 31 and 98 each, dominating the remaining observations. Removing the direct influence of these observations recovers a sensible point estimate; the bias correction we develop below allows this trimmed estimator to target the \emph{original} LATE with valid inference, rather than a redefined trimmed estimand. A natural first response is to trim observations with extreme propensity scores. Although intuitive, we highlight that trimming a DR estimator eliminates its double robustness, the very property that justified using DR methods in the first place.

We demonstrate this via simulations: when the outcome regression is misspecified but the propensity score is correct, the standard trimmed DR estimator should remain consistent, yet its 95\% coverage collapses to 34\% at a trimming threshold of $h = 0.10$ (Section~\ref{sec:trim_problem}). The double robustness guarantee is gone. The alternatives to trimming also come with important caveats. Changing the target parameter of interest \citep[e.g.,][]{CrumpEtAl2006_WP_MovingGoalposts, li2018balancing, Sloczynski2022_Restat,Sloczynski2026_Restud,Blandhol_etal_2026_Restud} addresses the variance problem by redefining the causal question being answered---paraphrasing the song \textit{When I'm Not Near the Girl I Love}, ``When I'm not near the parameter I love, I love the parameter I'm near.''\footnote{Our understanding is that Art Goldberger is the source of this paraphrase. We thank Matt Masten for bringing this to our attention.} A prominent example is imposing a linear specification and interpreting the regression treatment coefficient as a weighted average of heterogeneous treatment effects; the resulting estimand can have unattractive properties such as placing negative weight on some subpopulations and being hard to motivate via policy justification, and must be derived on a case-by-case (and design-by-design) basis \citep{Sloczynski2022_Restat, Sloczynski2026_Restud, Mogstad_Torgovitsky_2024, Caetano_Callaway_2024}.

This paper shows that robustness to weak covariate overlap need not come at the cost of redefining the causal estimand. We make three contributions. First, we establish a negative result: trimming without bias correction eliminates the double robustness property of DR estimators. The trimmed estimator is consistent only if the outcome regression is correctly specified, losing robustness to the very misspecification it was designed to guard against (Section~\ref{sec:trim_problem}). Second, we propose a class of bias-corrected DR estimators that is also robust against weak overlap. We show that a key condition for bias correction---that the conditional mean of the quantity being averaged vanishes as the propensity score approaches its boundary---holds under \emph{either} route to double robustness. When the propensity score is correctly specified, no observations of the relevant type exist at the boundary; when the outcome regression is correctly specified, the conditional expectation of the DR residual is zero everywhere. Either way, the conditional mean of the ratio has a well-defined limit at the boundary, and a low-dimensional polynomial can reconstruct the contribution of the trimmed observations, so the bias correction preserves double robustness. Third, we highlight that our unified framework covers the ATE, LATE, and staggered DiD designs and derive a closed-form influence function that enables standard inference. Implementation adds a single polynomial regression step to any existing DR procedure.

Rich conditioning sets are often needed for identification---interviewer fixed effects for the test score gap, county characteristics for parallel trends, income controls for instrument validity---but the same covariates that make identification credible can make estimation infeasible through weak overlap \citep{damour2021overlap}. Our estimator addresses this tension: researchers can condition on whatever covariates identification requires while substantially reducing the variance penalty. Four empirical applications---two DiD, one under unconfoundedness, and one using instrumental variables---demonstrate that the proposed estimator produces tighter confidence intervals and recovers informative estimates: in a DiD analysis of Medicaid expansion, the 95\% simultaneous confidence bands narrow from roughly 68 to 17 deaths per 100,000 on average over the post-treatment horizons; in an IV analysis of 401(k) participation, the estimator recovers a positive and significant savings effect of \$8,864, consistent with prior findings; and under unconfoundedness, it detects the Black--White test score gap at age two more precisely than linear regressions while directly targeting the ATE without imposing linearity. Across the four applications, standard errors from existing DR methods are 1.8 to 6.6 times larger than those from our proposed doubly robust bias-corrected (DR-BC) method. Since reducing the standard error by a factor of $c$ translates into a $c^2$-fold reduction in the sample size needed to detect a given effect at conventional power, these gains expand what applied researchers can learn from existing datasets.

Our approach builds on and extends several strands of work on causal inference under weak overlap \citep[e.g.,][]{crump2009dealing,khan/tamer:2010}.\footnote{See \citet{sasaki2018estimation} for more discussions on this literature.} 
We are mostly related to \citet{yang2018asymptotic} and \citet{heiler2019valid}, who study DR-type estimators in unconfoundedness settings; however, \citet{yang2018asymptotic} relies on trimming strategies that redefine the estimand, while \citet{heiler2019valid} does not consider trimming at all. None of these papers considers more general research designs as we do. \citet{sasaki2018estimation} propose bias-corrected inverse probability weighting estimators for a single ratio moment under weak overlap, relying on the boundary condition when the propensity score is correctly specified. Our framework differs in that it accommodates doubly robust estimands, establishes boundary conditions under both DR routes, and handles the nonlinear multi-moment structure required for the ATE, LATE, and DiD designs. In particular, the H\'ajek-type normalization that is standard in causal inference creates a denominator dependence across moments---requiring us to jointly bias-correct the normalization moments---and the residual-versus-normalization moment asymmetry we exploit under the outcome-correct route has no analog in the unnormalized, single-moment setting of \citet{sasaki2018estimation}. See, e.g., \citet{sloczynski2024abadie} for a recent discussion of why H\'ajek-type estimators are predominant in causal inference problems.

\medskip
\noindent\textbf{Notation.} We write $\E[\cdot]$ for the expectation operator and $\expen{\cdot}=n^{-1}\sum_{i=1}^n(\cdot)_i$ for the sample average. The indicator function is $\idx{\cdot}$. For a nuisance parameter $\gamma$, we let $\gamma_0$ denote the true value, $\hat\gamma$ an estimator, and $\gamma^*$ the probability limit of $\hat\gamma$.

\section{Overview and practical relevance}\label{sec:overview}

This section introduces the setup, demonstrates that trimming compromises double robustness, presents four empirical applications, and describes the proposed estimator.

\subsection{Setup and DR estimands}\label{sec:setup}

Our framework covers different research designs commonly used in empirical work, including unconfoundedness (selection on observables), LATE (instrumental variables), and DiD. Although the notation and identification assumptions differ across designs, they share a common structure: each involves a propensity score that can take values near its boundary, leading to weak overlap, and each admits a DR estimand that combines an outcome regression with inverse probability weighting. We present the DR estimands for these three designs using normalized weights that sum to one within each treatment group.

\subsubsection{Unconfoundedness}
A common goal in observational studies is to estimate the average effect of a binary treatment on an outcome, while adjusting for pre-treatment covariates to make the identification assumptions more plausible. Let $Y(1)$ and $Y(0)$ respectively denote the potential outcomes with and without treatment, $D \in \{0,1\}$ the treatment indicator, and $X$ a vector of pre-treatment confounding variables. The propensity score is $p_D(X) = \E[D \mid X]$, and the outcome regression is $m_Y(d, X) = \E[Y \mid D = d, X]$. Under unconfoundedness ($\{Y(0), Y(1)\} \indep D \mid X$) and overlap ($0 < p_D(X) < 1$), the $\text{ATE} = \E[Y(1) - Y(0)]$ is identified and admits the following DR representation \citep[see, e.g.,][and references therein]{Sloczynski2018}
\begin{equation}\label{eq:dr_ate}
\resizebox{0.92\linewidth}{!}{$\displaystyle
	\text{ATE}_{\text{DR}} = \E\bracks{m_Y(1,X)-m_Y(0,X)+w_{D=1}^{\text{ATE}}(D,X)\parens{Y - m_Y(1,X)} - w_{D=0}^{\text{ATE}}(D,X)\parens{Y - m_Y(0,X)}},
    $}
\end{equation}
where
\begin{align}\label{eq:weights_ate}
	w_{D=1}^{\text{ATE}}(D,X) = \frac{D/p_D(X)}{\E\bracks{D/p_D(X)}}, \quad w_{D=0}^{\text{ATE}}(D,X) = \frac{(1-D)/(1-p_D(X))}{\E\bracks{(1-D)/(1-p_D(X))}}.
\end{align}
When $p_D(X)$ is close to zero or one for some values of $X$, the weights in \eqref{eq:weights_ate} can become large. This is the weak overlap problem. Standard two-step estimators for \eqref{eq:dr_ate} do not achieve the $\sqrt{n}$-consistency in such weak-overlap cases \citep{khan/tamer:2010}.

\subsubsection{Instrumental variables and LATE}
When treatment is endogenous, and unconfoundedness is not empirically plausible, researchers often use an instrument to identify the average treatment effect among units whose treatment status is shifted by the instrument, i.e., the compliers \citep{imbens/angrist:1994}. Consider a setting with a binary instrument $Z$, binary treatment $D$, and observed covariates $X$, with potential treatments $D(1)$ and $D(0)$ and potential outcomes $Y(1), Y(0)$. Define the instrument propensity score as $p_Z(X) = \E[Z \mid X]$, the reduced-form and first-stage regressions as $m_{Y}^{\text{LATE}}(z, X) = \E[Y \mid Z = z, X]$ and $m_{D}^{\text{LATE}}(z, X) = \E[D \mid Z = z, X]$, respectively. Under the standard IV-LATE assumptions, the local average treatment effect $\text{LATE} = \E[Y(1) - Y(0) \mid D(1) > D(0)]$ is identified and the DR estimand for the LATE \citep{Belloni2017, Sloczynski2022, sloczynski2024abadie} can be written as
\begin{equation}\label{eq:dr_late}
\resizebox{0.92\linewidth}{!}{$\displaystyle
\text{LATE}_{\text{DR}} =
\frac{\E\bracks{m_Y^{\text{LATE}}(1,X)-m_Y^{\text{LATE}}(0,X)+w_{Z=1}^{\text{LATE}}(Z,X)\parens{Y - m_Y^{\text{LATE}}(1,X)} - w_{Z=0}^{\text{LATE}}(Z,X)\parens{Y - m_Y^{\text{LATE}}(0,X)}}}
{\E\bracks{m_D^{\text{LATE}}(1,X)-m_D^{\text{LATE}}(0,X)+w_{Z=1}^{\text{LATE}}(Z,X)\parens{D - m_D^{\text{LATE}}(1,X)} - w_{Z=0}^{\text{LATE}}(Z,X)\parens{D - m_D^{\text{LATE}}(0,X)}}}
$}
\end{equation}
where
\begin{align*}
	w_{Z=1}^{\text{LATE}}(Z,X) = \frac{Z/p_Z(X)}{\E\bracks{Z/p_Z(X)}}, \quad w_{Z=0}^{\text{LATE}}(Z,X) = \frac{(1-Z)/(1-p_Z(X))}{\E\bracks{(1-Z)/(1-p_Z(X))}}.
\end{align*}
The numerator identifies the reduced-form effect and the denominator the compliance share, so the ratio recovers the LATE. Weak overlap arises when $p_Z(X)$ is close to zero or one, potentially implying a high variance of two-step estimators based on \eqref{eq:dr_late}.

\subsubsection{Difference-in-Differences}
Another widely used research design that accommodates selection on unobservables is the DiD design. Consider a staggered DiD panel data setup with $T$ time periods \citep{Callaway2021}. Let $G$ denote the period of first treatment ($G = \infty$ if never treated), $X$ be a vector of pre-treatment covariates, and let $Y_t(g)$ be the potential outcome in period $t$ if a unit was first treated in period $g$. Let $\delta \geq 0$ denote the number of anticipation periods. Under limited anticipation and a conditional parallel trends assumption \citep{Callaway2021}, the group-time average treatment effect $\text{ATT}(g,t) = \E[Y_t(g) - Y_t(\infty) \mid G = g]$ is identified for $t \geq g - \delta$. The more aggregated event-study estimand
\begin{align}\label{eq:es}
	\text{ES}(e) = \sum_{g} w_{g,e}^{\text{es}} \cdot \text{ATT}(g, g+e),
\end{align}
where $w_{g,e}^{\text{es}} = \P(G = g \mid G + e \leq T, G \leq T)$ is also identified. Estimation of each $\text{ATT}(g,t)$ relies on a comparison group $\mathcal{C}_{g,t}$ (either never-treated or not-yet-treated units) with indicator $C_{g,t}$, generalized propensity score $p_{g,t}(X) = \P(G = g \mid X, \idx{G=g} + C_{g,t} = 1)$, and outcome regression $m_{g,t}(X) = \E[Y_t - Y_{g-\delta-1} \mid X, C_{g,t} = 1]$. The DR estimand for the $\text{ATT}(g,t)$ parameters is given by
\begin{align}\label{eq:dr_did}
	\text{ATT}_{\text{DR}}(g,t) = \E\bracks{\parens{w_{G=g}^{\text{DiD}}(G) - w_{g,t}^{\text{DiD}}(G,X)}\parens{Y_t-Y_{g-\delta-1} -m_{g,t}(X)}},
\end{align}
with weights given by
\begin{align*}
	w_{G=g}^{\text{DiD}}(G) = \frac{\idx{G=g}}{\E\bracks{\idx{G=g}}}, \quad w_{g,t}^{\text{DiD}}(G,X) = \frac{C_{g,t}\,p_{g,t}(X)/(1-p_{g,t}(X))}{\E\bracks{C_{g,t}\,p_{g,t}(X)/(1-p_{g,t}(X))}}.
\end{align*}
DR estimands for $\text{ES}(e)$ are defined by replacing $\text{ATT}(g, g+e)$ in \eqref{eq:es} with $\text{ATT}_{\text{DR}}(g,g+e)$ as defined in \eqref{eq:dr_did}. Weak overlap arises when $p_{g,t}(X)$ is close to one, potentially implying a high variance of DiD and ES estimators based on \eqref{eq:dr_did}.

\subsubsection{The common challenge}
Across all three designs, the DR estimand involves weights that divide by either the propensity score or one minus the propensity score (for the ATE, both treated and control weights can be extreme; for the LATE, both instrument groups; for DiD, only the comparison-group weights). When any such denominator is near zero, the resulting weights become very large. A handful of such observations can dominate the entire estimate. Trimming them reduces variance but introduces bias for the original estimand. As we show next, trimming a standard DR estimator can \emph{eliminate its double robustness property}.

\subsection{Trimming compromises double robustness}\label{sec:trim_problem}

A key appeal of the DR estimands presented above is their robustness to model misspecification: estimators based on them remain consistent for the causal parameter of interest even if one of the two working models (the propensity score or the outcome regression) is misspecified. A natural question is: Does this property survive when one trims observations with extreme propensity scores? The answer is no, as we illustrate via a simulation exercise.

We illustrate this with a DiD setup similar to \citet{Kang2007} and \citet{sant2020doubly}, with $n = 10{,}000$, a true ATT of zero, and weak overlap by design. To assess double robustness, we consider two complementary configurations: the propensity score is misspecified while the outcome regression is correct, and vice versa. We repeat each exercise $10{,}000$ times; full details of the data-generating process are in Appendix~\ref{appendix:simulation_dgp}, which also reports results for additional configurations (including correct specification of both models)---all yield qualitatively similar conclusions.

\begin{figure}[htp!]
	\centering
    \resizebox{0.93\textwidth}{!}{%
    \begin{minipage}{\textwidth}
		\begin{subfigure}[t]{.48\textwidth}
			\centering
			\includegraphics[width = \textwidth]{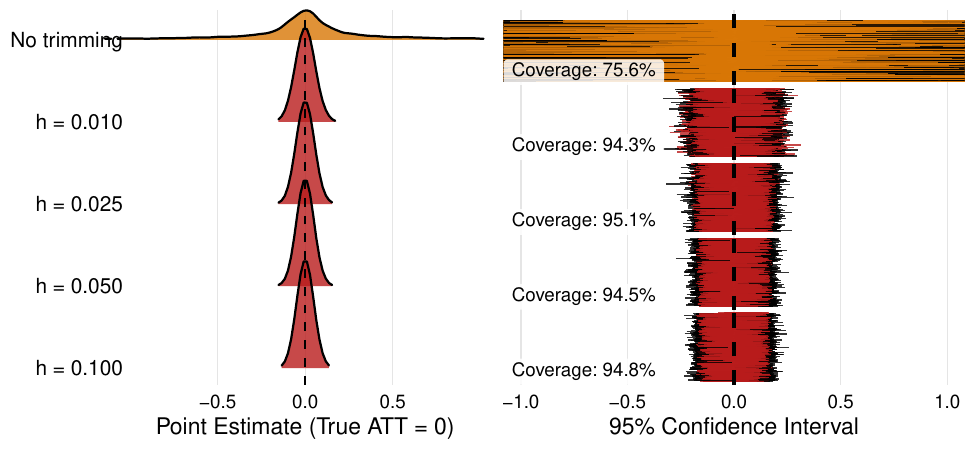}
			\caption{PS misspecified -- Standard trimmed DR}
			\label{fig:sim_trim_dgp2}
		\end{subfigure}
		\hfill
		\begin{subfigure}[t]{.48\textwidth}
			\centering
			\includegraphics[width = \textwidth]{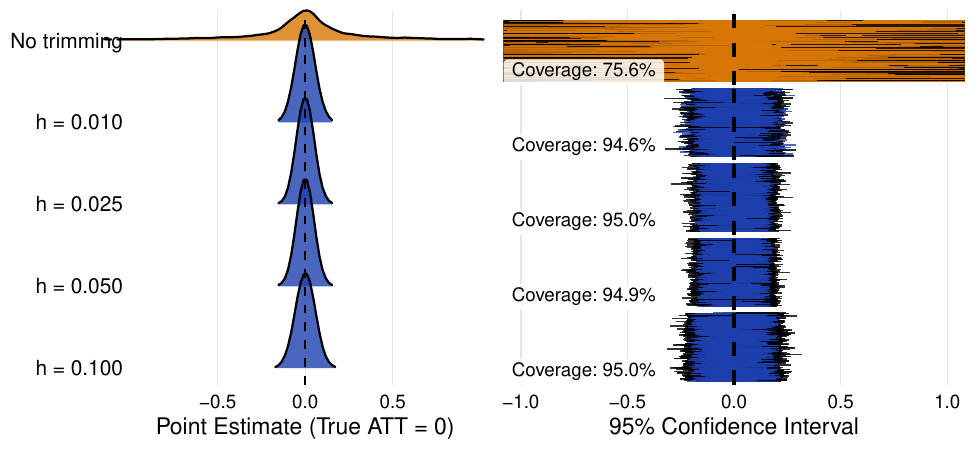}
			\caption{PS misspecified -- DR-BC (our proposal)}
			\label{fig:sim_bc_dgp2}
		\end{subfigure}\\[0.3cm]
		\begin{subfigure}[t]{.48\textwidth}
			\centering
            \includegraphics[width = \textwidth]{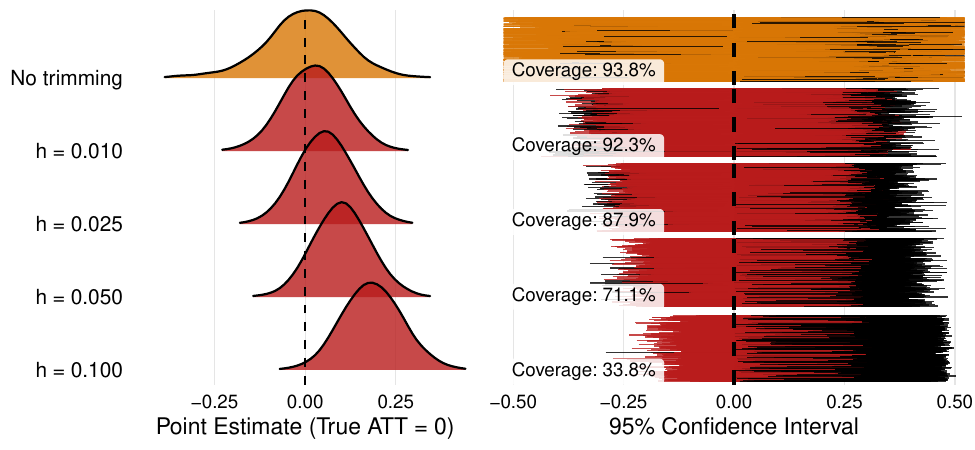}
            \caption{Outcome misspecified -- Standard trimmed DR}
            \label{fig:sim_trim_dgp3}
		\end{subfigure}
		\hfill
		\begin{subfigure}[t]{.48\textwidth}
			\centering
            \includegraphics[width = \textwidth]{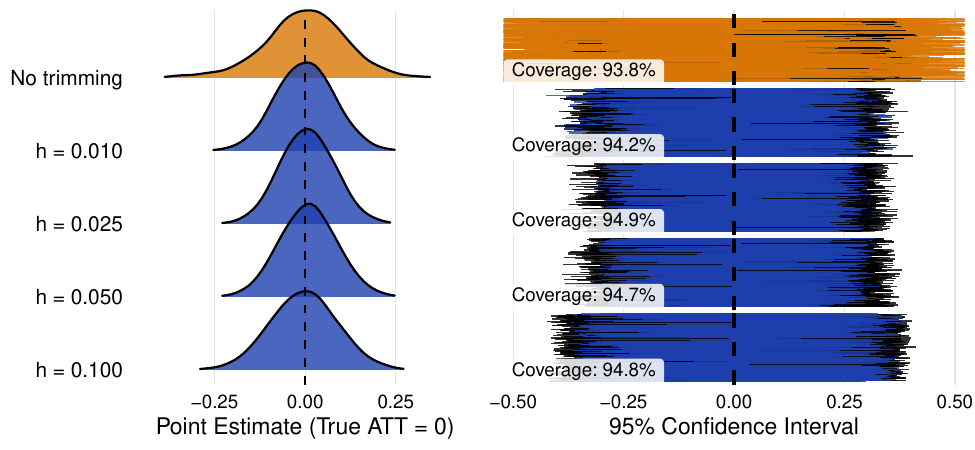}
            \caption{Outcome misspecified -- DR-BC (our proposal)}
            \label{fig:sim_bc_dgp3}
		\end{subfigure}
        \end{minipage}
        }
		\vspace{-0.2cm}
		\caption{Effect of trimming on DR DiD estimators under model misspecification}
		\label{fig:Sims_DiD}
        \justifying
	\vspace{0.1cm}\noindent\scriptsize{\textit{Notes.} Simulation design as discussed in Appendix \ref{appendix:simulation_dgp}, with $n=10{,}000$ and $10{,}000$ Monte Carlo repetitions. The true ATT is zero. Top row (Panels~(a)--(b)): propensity score misspecified, outcome regression correct. Bottom row (Panels~(c)--(d)): propensity score correct, outcome regression misspecified. In both cases, double robustness should ensure consistent estimation. Left column (Panels~(a),~(c)): standard trimmed DR estimator. Right column (Panels~(b),~(d)): our bias-corrected estimator (DR-BC). Ridge plots show the density of point estimates; zipper plots show 95\% confidence intervals across $10{,}000$ Monte Carlo draws, with black segments indicating non-coverage.}
\end{figure}

We use the standard DR DiD estimator based on \eqref{eq:dr_did}, with logit for the propensity score and OLS for the outcome regression. We vary the trimming threshold over $h \in \{0, 0.01, 0.025, 0.05, 0.10\}$, where $h=0$ is the no-trimming case. Figure~\ref{fig:Sims_DiD} summarizes the results, showing ridge density plots alongside zipper confidence interval plots (black segments indicate non-coverage).

The left column of Figure~\ref{fig:Sims_DiD} presents the results for the standard trimmed DR DiD estimator. When the propensity score is misspecified but the outcome regression is correct (Panel~(a)), the trimmed estimator performs well: bias is negligible and coverage remains near 95\% across all positive thresholds, as one would expect from a doubly robust estimator. However, the reverse case (Panel~(c)) reveals the problem: when the outcome regression is misspecified and the propensity score is correct, trimming introduces growing bias. At $h = 0.05$, the 95\% empirical coverage drops to 71\%, and at $h = 0.10$, it collapses to 34\%. The double robustness property is lost, as the reliability of this procedure depends on the outcome model being correctly specified. We expand on this rationale in Section~\ref{sec:dr_bc}.

The right column of Figure~\ref{fig:Sims_DiD} (Panels~(b) and~(d)) shows that our proposed weak-overlap-robust DR estimator resolves this trade-off. It trims comparison-group observations with extreme propensity scores, just as the standard trimmed estimator does, but adds a bias-correction term that compensates for the trimmed observations. In both configurations, whether the propensity score or the outcome regression is the correctly specified component, the estimator remains well-centered at the truth, and coverage stays near the nominal 95\% level (94--95\%). Confidence intervals at $h = 0.05$ are narrower than at $h = 0$, so the bias-corrected estimator retains double robustness while achieving the variance reduction that trimming provides. Thus, although the standard trimming procedure compromises the DR property, our proposed bias-correction procedure restores it and yields estimators with attractive statistical guarantees.


\subsection{Empirical applications}\label{sec:applications}
We compare DR-BC ($h = 0.05$) with the standard untrimmed DR estimator ($h = 0$) in four applications spanning all three research designs: unconfoundedness, IV and DiD.  Because both estimators target the same causal parameter, differences in standard errors reflect genuine precision gains, not changes in what is being estimated or the causal question being addressed.\footnote{In all applications, propensity scores are estimated via logit regression and outcome regressions via least squares. Standard errors are computed analytically from the influence function. For the DiD event studies, standard errors are clustered at the county level (Medicaid) and municipality level (Dias--Fontes), and simultaneous confidence bands are based on multiplier bootstrap.}
Because we focus on applications where weak overlap is a concern, the precision gains below are large; in settings with adequate overlap, the gains would be more modest.

\subsubsection{Difference-in-Differences}\label{sec:app_did}

We apply our method to two staggered DiD settings using the DR framework of \citet{Callaway2021}. In both cases, a rich set of covariates is used to make conditional parallel trends more plausible, but conditioning on them leads to weak overlap.

\begin{figure}[h!t]
	\centering
	\resizebox{0.7\textwidth}{!}{%
	\begin{minipage}{\textwidth}
		\begin{subfigure}[t]{\textwidth}
			\centering
			\includegraphics[width = \textwidth]{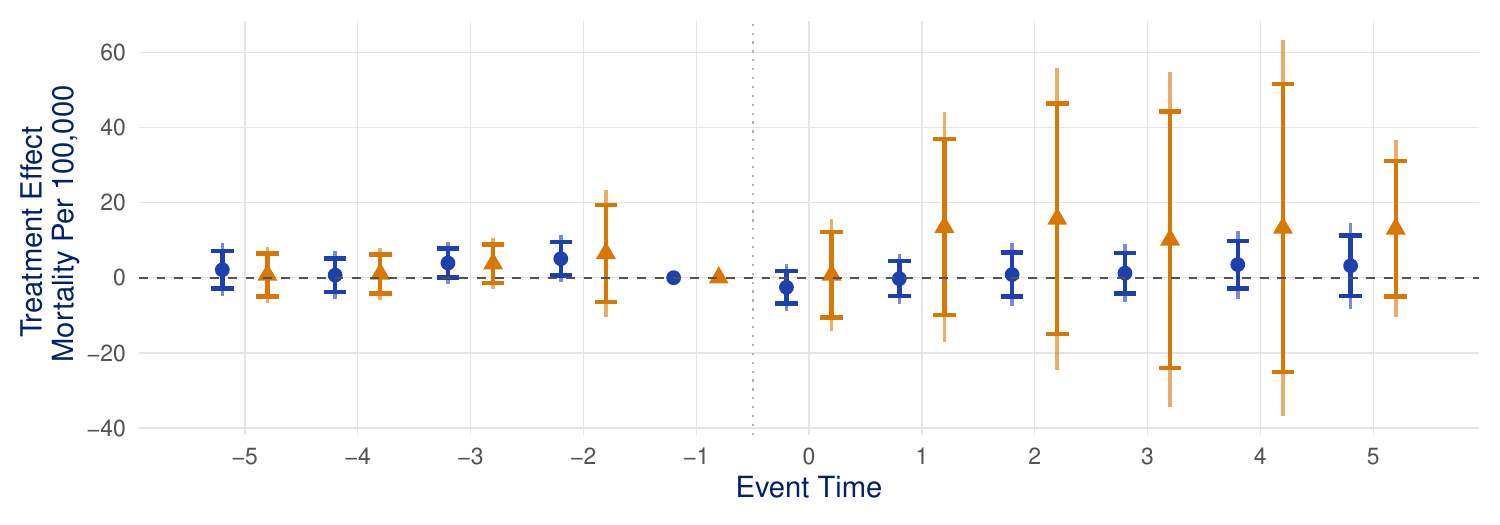}
		\vspace{-0.3cm}\caption{Medicaid expansion on mortality}
\label{fig:DiD_Medicaid}
		\end{subfigure}
		\vspace{0.1cm}
		\begin{subfigure}[t]{\textwidth}
			\centering
			\includegraphics[width = \textwidth]{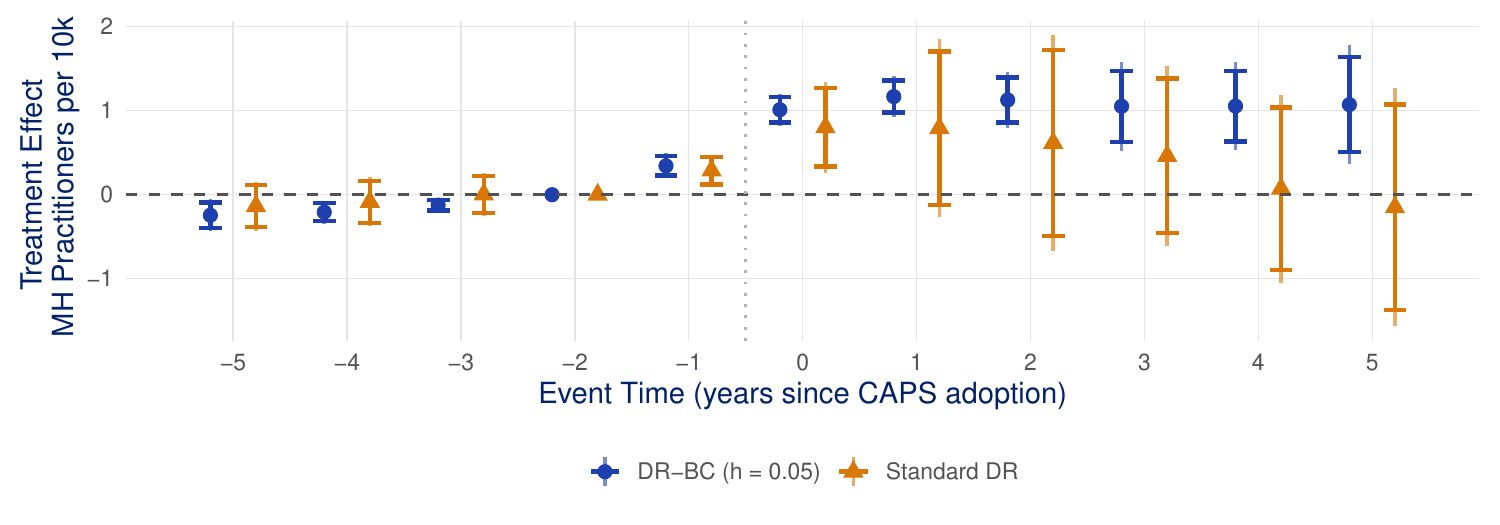}
		\vspace{-0.3cm}\caption{CAPS adoption on mental health practitioners}
\label{fig:DiD_Dias_Fontes}
		\end{subfigure}
	\end{minipage}%
	}
          \caption{DR DiD event-study estimators across applications}
    \label{fig:DiD_applications}
        \justifying
	\vspace{0.1cm}\noindent\scriptsize{\textit{Notes.} Panel~(a) shows population-weighted event-study estimates with staggered treatment timing using the doubly robust estimation method of \citet{Callaway2021}. The comparison group consists of counties that remained untreated by 2019. The outcome is the crude mortality rate for adults ages 20--64. Data are from \citet{baker2025difference}. Panel~(b) shows event-study estimates with not-yet-treated municipalities as the comparison group. The outcome is mental health practitioners per 10,000 population. Covariates include state fixed effects and 29 baseline characteristics. Three small northern states with fewer than 10 treated municipalities are excluded. We allow for a one-year anticipation. Data are from \citet{DiasFontes2024_AEJPolicy}. In both panels, DR-BC is our proposed estimator with trimming threshold $h = 0.05$; Standard DR is the untrimmed estimator of \citet{Callaway2021}. Points denote point estimates, vertical bars denote 95\% simultaneous confidence bands, and error bars with caps denote 95\% pointwise confidence intervals.
   }
\end{figure}

\paragraph{Medicaid expansion and mortality.} Does Medicaid expansion save lives? \citet{baker2025difference} provides a DiD practical guide and uses this question as their running example, using county-level all-cause mortality rates for adults ages 20--64 and the staggered adoption of Medicaid expansion across states from 2009 to 2019. We re-analyze their data using the DR DiD framework of \citet{Callaway2021} with never-treated counties as the comparison group, conditioning on six county-level covariates (poverty rate, unemployment rate, median household income, percent female, percent White, and percent Hispanic) and using population weights. Because these covariates differ between expanding and non-expanding states, the generalized propensity score is close to 1 for some comparison counties, resulting in weak overlap; at $h = 0.05$, up to 3.3\% of comparison-group observations are trimmed in the most affected group-time cell. Panel~(a) of Figure~\ref{fig:DiD_applications} shows the event-study estimates. Post-treatment standard errors are on average 3.5 times larger for the standard DR estimator. While neither estimator rejects a zero effect, the standard DR 95\% simultaneous confidence bands span on average 68 deaths per 100,000, which is too wide to be policy relevant. DR-BC narrows these bands to roughly 17 deaths per 100,000 on average, ruling out large effects in either direction and providing a much more informative bound on the plausible magnitude of Medicaid expansion's mortality impact. 

\paragraph{Mental health reform in Brazil.} Did Brazil's 2002 psychiatric reform expand community mental health capacity? \citet{DiasFontes2024_AEJPolicy} study this question using the staggered rollout of community-based Psychosocial Care Centers (CAPS) across municipalities, which replaced centralized hospital-based care. Because the timing of CAPS adoption was correlated with municipal characteristics, covariate adjustment is essential for the parallel trends argument. We re-analyze their data using the DR DiD framework of \citet{Callaway2021}, focusing on mental health practitioners per 10,000 population; results for additional outcomes (service utilization, hospital spending, violence) are in Appendix \ref{appendix:did_outcomes}. We restrict the sample to states with at least 10 treated municipalities and adjust for state fixed effects and 29 baseline covariates that are motivated by the original study. Because this rich covariate set leaves limited overlap between treated and not-yet-treated municipalities, a small number of comparison observations receive estimated propensity scores near one and therefore extremely large weights; at $h = 0.05$, up to 0.2\% of comparison-group observations are trimmed in the most affected group-time cell. Panel~(b) of Figure~\ref{fig:DiD_applications} presents the event-study estimates. Post-treatment standard errors are on average 2.8 times larger for the standard DR estimator, and at longer horizons the standard DR confidence intervals become too wide to be informative. DR-BC yields estimates precise enough to trace the workforce expansion over time, with a sustained increase of roughly 1 mental health practitioner per 10,000 population following CAPS adoption, consistent with the findings of \citet{DiasFontes2024_AEJPolicy}. See Appendix \ref{appendix:did_outcomes} and Appendix \ref{appendix:h_sensitivity} for additional outcomes and sensitivity analysis.

\subsubsection{Unconfoundedness}\label{sec:app_ate}

How large is the Black--White test score gap at age two, and can we estimate it without imposing functional form restrictions? Using data from the Early Childhood Longitudinal Study, Birth Cohort (ECLS-B),
\citet{FryerLevitt2013_AER} document that the gap is negligible at 9 months but grows to roughly 0.2--0.4 standard deviations by age two, conditioning on a rich set of family background variables via OLS. Because their OLS specification imposes linearity and does not interact race dummies with covariates, it may not recover the ATE under treatment effect heterogeneity \citep{Sloczynski2022_Restat}. We re-analyze their data using DR estimation, treating race as the ``treatment'' in an operational sense: following \citet{holland1986}, the ``ATE'' here is interpreted as a covariate-adjusted descriptive contrast between Black and White children. Their conditioning set includes interviewer fixed effects, fifth-degree polynomials in parental age, and numerous interactions. The interviewer fixed effects create many small cells --- interviewers who observed children of mostly one race generate near-zero or near-one propensity scores --- pushing the propensity score to its boundaries for a non-negligible share of the sample. At $h = 0.05$, a small number of observations with extreme arm-specific weights drive the instability in the untrimmed DR estimator (see Appendix~\ref{appendix:ps_density} for the exact decomposition and Appendix~\ref{appendix:h_sensitivity} for sensitivity analysis).

\begin{figure}[htbp]
	\centering
	\resizebox{0.7\textwidth}{!}{%
	\begin{minipage}{\textwidth}
		\centering
		\includegraphics[width = \textwidth]{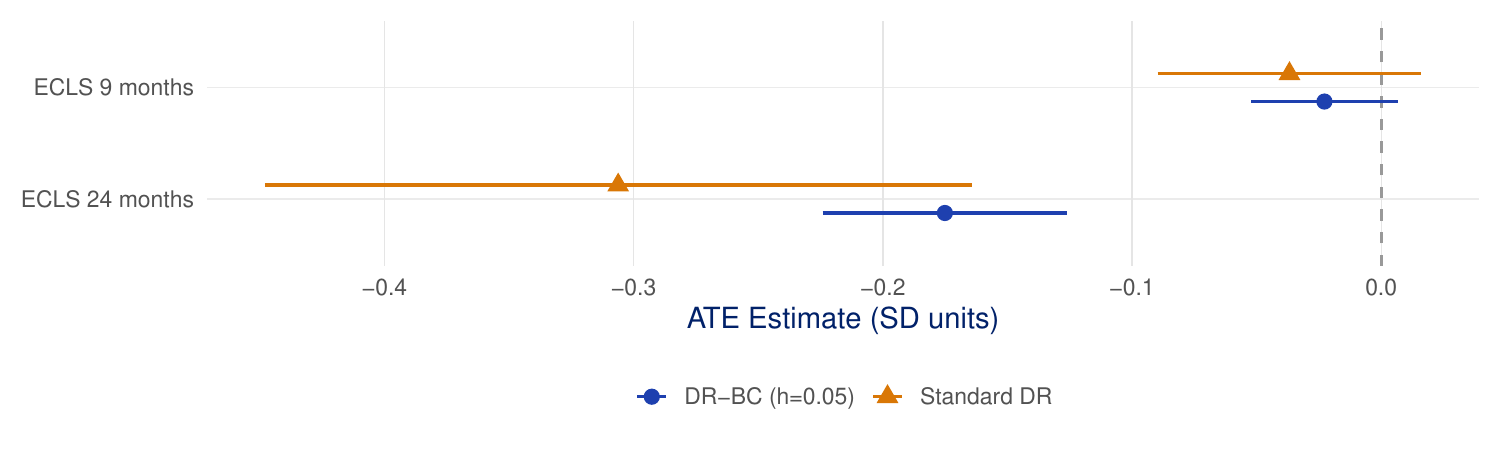}
	\end{minipage}%
	}
          \caption{DR estimates of the Black--White test score gap}
    \label{fig:Fryer_Levitt}
        \justifying
	\vspace{0.1cm}\noindent\scriptsize{\textit{Notes.} Forest plot of ATE estimates for the Black--White test score gap at ages 9 months and 24 months from the ECLS-B. DR-BC is our proposed estimator with $h = 0.05$; Standard DR is the untrimmed estimator. Covariates and sample restrictions follow \citet{FryerLevitt2013_AER}. Error bars denote 95\% confidence intervals based on analytical standard errors.
   }
\end{figure}

Figure~\ref{fig:Fryer_Levitt} presents the results. At 24 months, the standard error drops from 0.072 to 0.025 (a 2.9-fold reduction), while at 9 months the reduction is 1.8-fold (from 0.027 to 0.015). DR-BC yields $-0.175$ (SE $= 0.025$) under a much more flexible specification that does not impose linearity, while the standard DR estimate of $-0.306$ (SE $= 0.072$) illustrates the precision cost of flexibility without overlap correction. DR-BC thus delivers OLS-level precision while directly targeting the ATE, without strong functional-form restrictions; the $h$-sensitivity analysis in Appendix \ref{appendix:h_sensitivity} (Figure~\ref{fig:h_sensitivity_ATE}) confirms that point estimates remain stable across $h$.

\subsubsection{Instrumental variables, LATE, and ITT}\label{sec:app_late}

Does 401(k) participation increase household savings? A central challenge is that participation is endogenous. A widely used identification strategy exploits employer-determined \emph{eligibility} as an instrument for participation \citep{abadie2003semiparametric, Chernozhukov2004, Benjamin2003a}. Using the 1991 Survey of Income and Program Participation (SIPP) data and the same covariates as \citet{abadie2003semiparametric}, we estimate both the ITT effect of eligibility and the LATE of participation for two outcomes: net financial assets and total wealth. We consider two samples: a full sample requiring only positive income \citep{Chernozhukov2004, Benjamin2003a}, and a restricted sample with income between \$10,000 and \$200,000 \citep{abadie2003semiparametric}. The income restriction was motivated precisely by weak overlap: as \citet[p.~249]{abadie2003semiparametric} notes, ``outside this interval, 401(k) eligibility in the sample is rare.'' Indeed, there are a few variance-inflating observations in this dataset: in the full sample, the 2 ineligible observations in the $\hat{p}_Z(X)>0.95$ tail have effective weights $1/(1-\hat{p}_Z(X))$ of roughly 31 and 98. It is this small set of high-weight ineligible observations that drives the variance inflation, yet removing their direct contribution via trimming is enough to reverse the sign of the untrimmed estimate. 

\begin{figure}[htbp]
	\centering
	\resizebox{0.8\textwidth}{!}{%
	\begin{minipage}{\textwidth}
		\begin{subfigure}[t]{.48\textwidth}
			\centering
			\includegraphics[width = \textwidth]{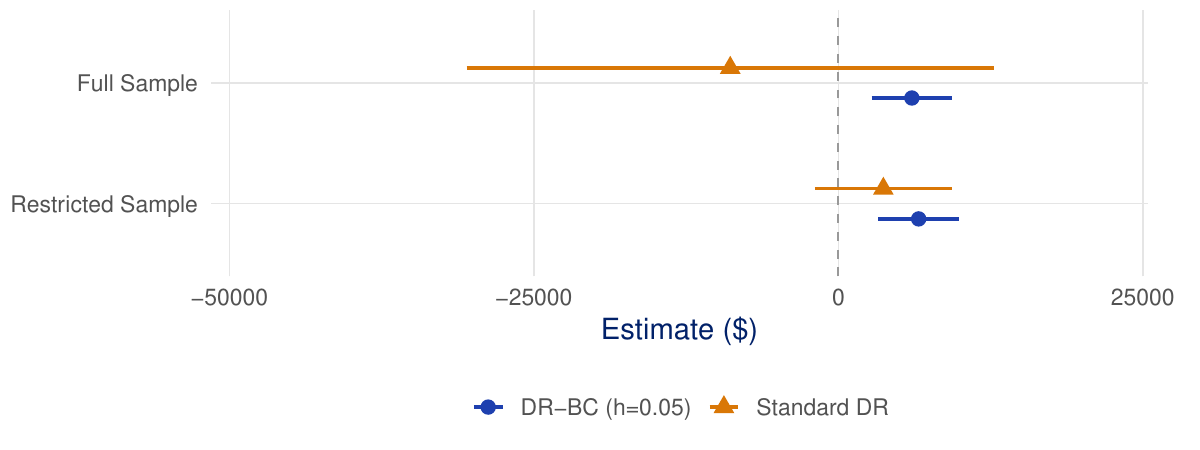}
			\caption{DR ITT estimates of having access to 401(k) on net financial assets}
			\label{fig:401K_itt_nfa}
		\end{subfigure}
		\hfill
		\begin{subfigure}[t]{.48\textwidth}
			\centering
			\includegraphics[width = \textwidth]{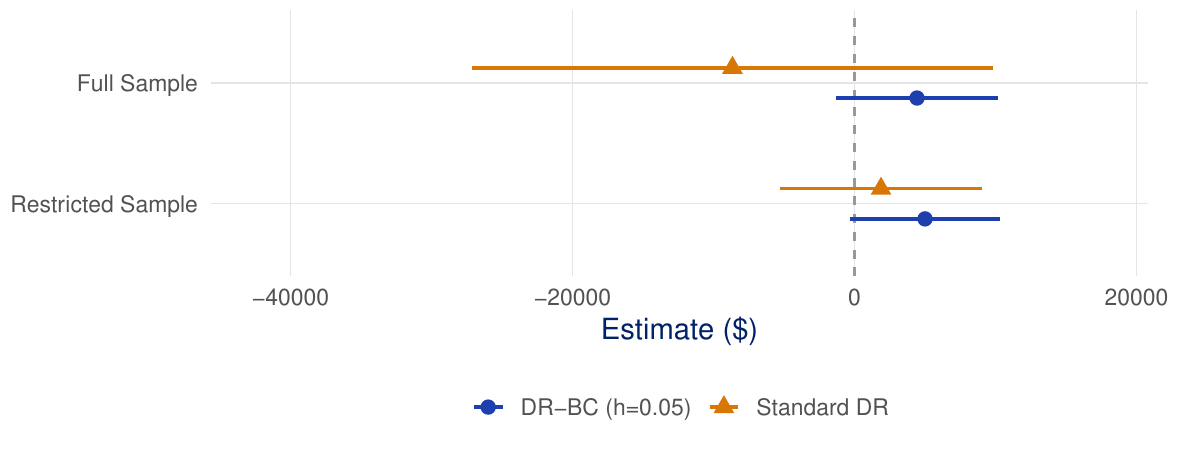}
			\caption{DR ITT estimates of having access to 401(k) on total wealth}
			\label{fig:401K_itt_tw}
		\end{subfigure}\\[0.3cm]
		\begin{subfigure}[t]{.48\textwidth}
			\centering
			\includegraphics[width = \textwidth]{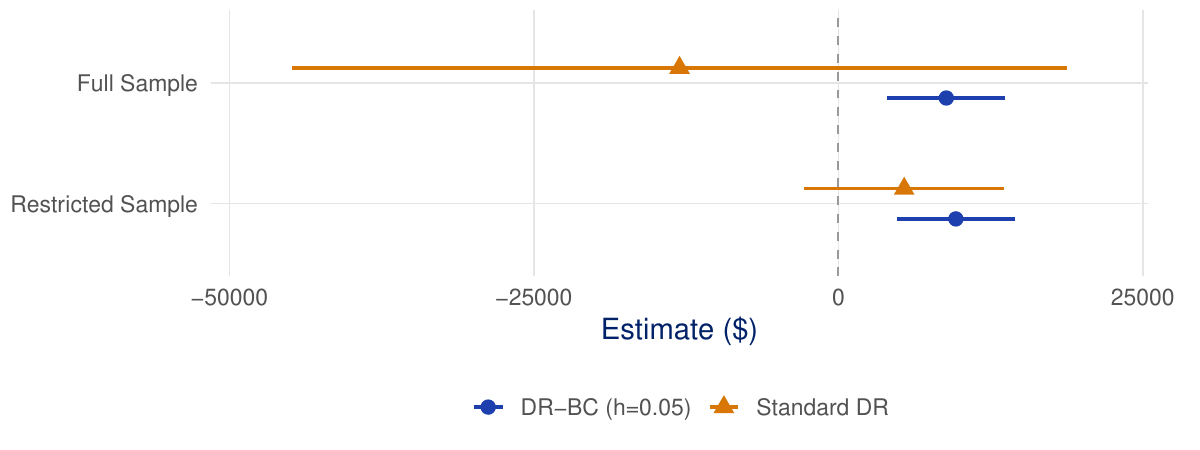}
			\caption{DR LATE estimates of 401(k) participation on net financial assets}
			\label{fig:401K_late_nfa}
		\end{subfigure}
		\hfill
		\begin{subfigure}[t]{.48\textwidth}
			\centering
			\includegraphics[width = \textwidth]{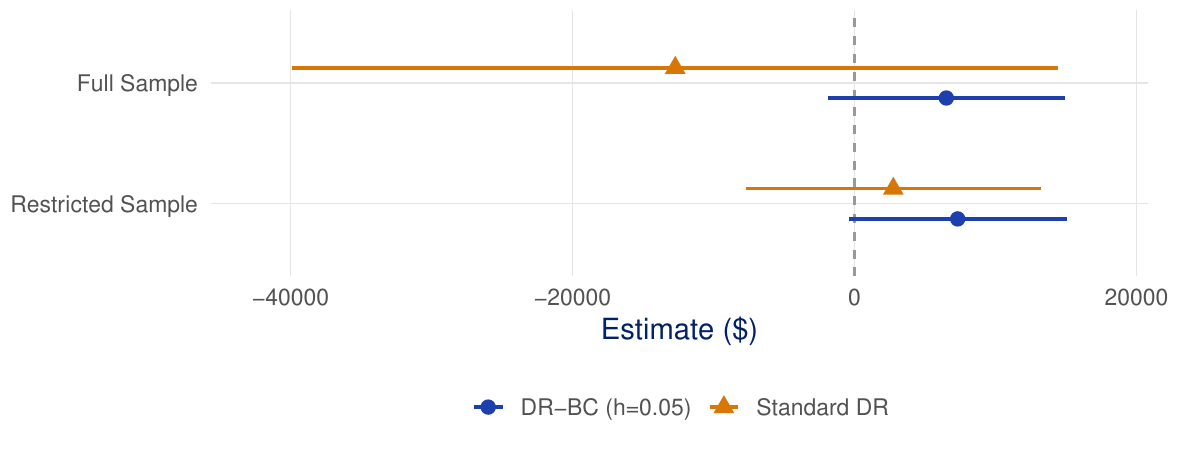}
			\caption{DR LATE estimates of 401(k) participation on total wealth}
			\label{fig:401K_late_tw}
		\end{subfigure}
	\end{minipage}%
	}
	\caption{Effect of 401(k) retirement plans on asset accumulation}
	\label{fig:401K}
        \justifying
	\vspace{0.1cm}\noindent\scriptsize{\textit{Notes.} Full sample restricts to households with positive income \citep{Benjamin2003a, Chernozhukov2004}. Restricted sample further restricts to income between \$10,000 and \$200,000 \citep{abadie2003semiparametric}. DR-BC trims observations with estimated instrument propensity scores $\widehat{p}_Z(X)$ outside $[0.05, 0.95]$; Standard DR is the untrimmed DR estimator based on \eqref{eq:dr_late}. Covariates: income, age, age squared, marital status, family size. Points denote estimates; horizontal lines denote 95\% confidence intervals.
   }
\end{figure}

Figure~\ref{fig:401K} presents the results. The full-sample LATE for net financial assets illustrates how weak overlap can render standard DR estimates unreliable: the standard DR point estimate is $-\$13{,}042$ (SE $= \$16{,}223$), suggesting that 401(k) participation \emph{reduces} savings---a conclusion that is at odds with standard economic theory and the existing literature. The untrimmed DR estimator is uninformative in this sample, with a confidence interval spanning from $-\$45{,}000$ to $+\$19{,}000$.

DR-BC yields $\$8{,}864$ (SE $= \$2{,}471$), recovering a positive, statistically significant effect consistent with prior findings \citep{abadie2003semiparametric}. The sign reversal is not a change in estimand---both estimators target the same LATE---but reflects removing the outsized influence of the two ineligible observations whose effective weights dominate the untrimmed estimator. Decomposing the Wald ratio reveals that the instability is entirely in the reduced form: the first-stage estimate is stable at $0.68$ regardless of trimming, but the intention-to-treat (ITT) swings from $-\$8{,}879$ (SE $= \$11{,}044$) to $\$6{,}035$ (SE $= \$1{,}686$). 

The precision gains here come primarily from variance reduction through trimming; the bias correction ensures theoretical integrity by targeting the original LATE and providing the correct influence function. DR-BC point estimates are stable across $h \in [0.03, 0.10]$ (Appendix~\ref{appendix:h_sensitivity}). For total wealth, the standard error reduction is 3.2-fold. In the restricted sample, gains are more modest (1.35- to 1.70-fold), but point estimates are stable across both samples.

\subsection{Trimming with bias correction to retain the DR property}\label{sec:dr_bc}


Each of the DR estimands in Section~\ref{sec:setup}, whether for the ATE, LATE, or DiD ATT, can be written as a known function of one or more ratio moments $\alpha_\ell \equiv \E\bracks{B_\ell / A_\ell}$, where $\ell\in\{1,\cdots,L\}$ indexes the ratio moments entering the estimand. Here, $A_\ell$ involves the propensity score and can be close to zero, and $B_\ell$ contains a group indicator---multiplied by an outcome residual for the effect-estimating moments, or standing alone for the companion normalization moments. The specific $(A_\ell, B_\ell)$ decomposition depends on the design, but the weak overlap problem is always the same: when $A_\ell$ is near zero, $B_\ell / A_\ell$ is unstable with high variance. Standard trimming discards observations with $|A_\ell| < h$ for some threshold $h > 0$. This stabilizes the estimator but changes what is being estimated:
\[
\E\bracks{\frac{B_\ell}{A_\ell} \cdot \idx{|A_\ell| \geq h}} \neq \E\bracks{\frac{B_\ell}{A_\ell}}.
\]
The trimming bias equals $\E\bracks{(B_\ell / A_\ell) \cdot \idx{|A_\ell| < h}}$, which does not vanish for fixed $h$. Our bias-correction term uses the behavior of the conditional expectation $\xi_\ell(a) \equiv \E\bracks{B_\ell \mid A_\ell = a}$ near the propensity score boundary to compensate for the discarded observations.

The key observation is that, under either condition for double robustness, the relevant conditional mean at the propensity score boundary equals zero. There are two routes. \emph{First}, when the propensity score is correctly specified, $A_\ell = 0$ places the propensity score at its boundary, so one group has probability zero at that covariate value. The group indicator in $B_\ell$ then forces $B_\ell = 0$ almost surely conditional on $A_\ell = 0$, giving $\xi_\ell(0) = 0$. \emph{Second}, when the outcome regression is correctly specified, the residual-containing moments have a conditional mean of zero given $X$. Since $A_\ell$ is a function of $X$, the law of iterated expectations gives $\xi_\ell(a)  = 0$ for those moments, and in particular $\xi_\ell(0) = 0$. The companion normalization moments need not satisfy this stronger statement, but they are irrelevant along this route because the corresponding residual moments are already zero. One can verify this for each estimand in Section~\ref{sec:setup}; the formal general statement is Assumption~\ref{a:m_function} in Section~\ref{sec:theory}.

Note the asymmetry: when the outcome regression is correct, $\xi_\ell(a) = 0$ for \emph{all}~$a$ for the residual-containing moments, so trimming introduces no bias and the correction is inoperative. It is needed only when the propensity score is correct and the outcome model is not. In either case, $\xi_\ell(0) = 0$ implies that $\xi_\ell(a)/a$ has a removable singularity at zero and can be approximated by a polynomial in $a$. This explains the results in Figure~\ref{fig:Sims_DiD}.

Our proposed estimator replaces each problematic ratio moment $\E\bracks{B_\ell / A_\ell}$ with its bias-corrected counterpart:
\begin{align*}
	\widehat{\alpha}_\ell(h) = \underbrace{\expen{\frac{B_\ell}{A_\ell} \cdot \idx{|A_\ell| \geq h}}}_{\text{trimmed mean}} + \underbrace{\sum_{\kappa=1}^{k} \frac{\expen{A_\ell^{\kappa-1} \cdot \idx{|A_\ell| < h}}}{\kappa!} \cdot \widehat{\xi}_\ell^{(\kappa)}(0)}_{\text{bias correction}},
\end{align*}
where $\widehat{\xi}_\ell^{(\kappa)}(0)$ is the $\kappa$-th derivative of a sieve estimator for $\xi_\ell(\cdot)$, evaluated at zero, using a shifted Legendre polynomial basis of maximum degree $K$, and $k$ is the order of the bias-correction terms. The first part is the standard trimmed estimator. The second part uses the observations excluded by trimming ($|A_\ell| < h$), together with the estimated derivatives, to reconstruct their contribution. The final estimator applies a known function $\Lambda$ to the corrected moments $(\widehat{\alpha}_1(h), \ldots, \widehat{\alpha}_L(h))$; with the normalized weights in Section~\ref{sec:setup}, $\Lambda$ involves ratios for all three designs.

This structure preserves double robustness. Under either DR condition, $\xi_\ell(0)=0$ holds for the residual-containing moments under correct outcome regression, and all moments under correct propensity score, so the polynomial approximation of $\xi_\ell(a)/a$ is valid for those moments. Under standard regularity conditions on the propensity score density and the smoothness of $\xi_\ell$, the resulting estimator is consistent and asymptotically normal for the original estimand, with a closed-form influence function that enables standard inference; the standard errors account for all sources of estimation uncertainty, including the sieve-based bias correction (Section~\ref{sec:theory}).

In practice, we recommend $h = 0.05$, $k = 1$ (first-order bias correction), and $K = 3$. Figure~\ref{fig:Sims_DiD} provides evidence: coverage remains near 95\% under both misspecification configurations, while confidence intervals narrow relative to $h = 0$. Higher-order corrections ($k \geq 2$) degrade finite-sample performance. Sensitivity to $h$ is examined in Appendix~\ref{appendix:h_sensitivity}. Section~\ref{sec:theory} presents the formal theory.

\subsection{Practical diagnostics and when to use DR-BC}\label{sec:diagnostics}

DR-BC and the standard (untrimmed) DR estimator target the same causal parameter, and DR-BC reduces to standard DR when no observations are trimmed ($h = 0$ or no propensity scores near the boundary). The cost of using DR-BC when overlap is adequate is therefore minimal: it adds a polynomial regression step that has a negligible impact on the estimates. The benefit when overlap is weak can be substantial, as the applications above demonstrate. When the number of units trimmed is fairly small, the bias correction has little to reconstruct, and the precision gains come primarily from variance reduction. We also recommend plotting estimates as a function of $h$ (Appendix~\ref{appendix:h_sensitivity}) to assess stability; instability may indicate that the conditions underlying our results are not met.

\section{Theory}\label{sec:theory}

This section formalizes the general estimator introduced in Section~\ref{sec:dr_bc}, states the conditions for asymptotic normality, and verifies them for each research design. 

\subsection{General framework}\label{sec:general_framework}

The DR estimands in Section~\ref{sec:setup} share a common structure. Each can be written as a known function of $L$ ratio moments:
\begin{equation*}
\theta_0=\Lambda\left(\E\bracks{\frac{B_1(\gamma_0)}{A_1(\gamma_0)}},\ldots,\E\bracks{\frac{B_L(\gamma_0)}{A_L(\gamma_0)}}\right),
\end{equation*}
where $(A_\ell(\gamma),B_\ell(\gamma))_{\ell=1}^L$ are known functions of observed data, $\gamma_0=(m,p)$ collects the true outcome regression $m$ and propensity score $p$, and $\Lambda$ is a known function.\footnote{$\Lambda$ may also depend on $\gamma$ through regression adjustments $\E\bracks{m(d,X)}$ that are smooth, root-$n$ estimable, and not subject to weak overlap; see Section~\ref{sec:verification}.} Weak overlap means some $A_\ell(\gamma_0)$ can be close to zero, making $B_\ell/A_\ell$ potentially unstable with high variance.

Given i.i.d.\ observations $W_1,\ldots,W_n$, a preliminary estimator $\hat\gamma=(\hat m,\hat p)$ with probability limit $\gamma^*=(m^*,p^*)$ (so $m^*=m$ when the outcome model is correctly specified, or $p^*=p$ when the propensity score model is correctly specified), a trimming threshold $h > 0$, and positive integers $k \leq K$, our estimator is $\hat\theta=\Lambda(\hat\alpha_1(h,\hat\gamma),\ldots,\hat\alpha_L(h,\hat\gamma))$, where
\begin{equation}\label{eq:alpha_hat}
\hat\alpha_\ell(h,\gamma)=\expen{\frac{B_\ell(\gamma)}{A_\ell(\gamma)}\idx{|A_\ell(\gamma)|\geq h}}+\sum_{\kappa=1}^{k}\frac{\expen{A_\ell(\gamma)^{\kappa-1}\idx{|A_\ell(\gamma)|< h}}}{\kappa !} \cdot \hat{\xi}_\ell^{(\kappa)}(0;\gamma).
\end{equation}
Here $\hat{\xi}_\ell^{(\kappa)}(0;\gamma)$ is the $\kappa$-th derivative at zero of a sieve estimator for ${\xi}_\ell(a;\gamma)\equiv\E\bracks{B_\ell(\gamma)\mid A_\ell(\gamma)=a}$, the conditional mean of $B_\ell$ given $A_\ell$. The dependence on $\gamma$ was suppressed in Section~\ref{sec:dr_bc}. The sieve uses a shifted Legendre polynomial basis $q_K(\cdot)$ of degree $K$, natural since propensity scores lie in $[0,1]$; with $K$ fixed, the sieve basis is well-conditioned. The asymptotic theory allows $K$ to grow with $n$, and we recommend $K=3$ as a practical default for typical sample sizes in applications. Details are in Appendix \ref{appendix:sieve}.

\subsection{Asymptotic theory}\label{sec:assumptions}

We decompose $\hat\theta-\theta_0=(\hat\theta-\theta_h)+(\theta_h-\theta_0)$, where
$\theta_h\equiv\theta_h(\gamma_0)$ with $\theta_h(\gamma)=\Lambda(\alpha_1(h,\gamma),\ldots,\alpha_L(h,\gamma))$
is the population analog of $\hat\theta$. The first term is the estimation error; the second is the trimming bias. We require the following assumptions.

\begin{assumption}[Approximate double robustness]\label{a:double_robustness}
  If either $m^*=m$ or $p^*=p$ (i.e., the outcome regression or propensity score is correctly specified), then
   $\theta_h(\gamma^*)=\theta_0+o(h^k)$.
\end{assumption}

\begin{assumption}[Conditional mean at zero]\label{a:m_function}
For each $\ell=1,\ldots,L$ with $0\in\mathrm{support}(A_\ell(\gamma^*))$: \textnormal{(i)}~${\xi}_\ell(0;\gamma^*)=0$ if $p^*=p$, and ${\xi}_\ell(a;\gamma^*)=0$ for all $a$ for the residual moments if $m^*=m$; \textnormal{(ii)}~${\xi}_\ell(\cdot;\gamma^*)$ is $(k+1)$-times continuously differentiable near~$0$.
\end{assumption}

\begin{assumption}[Smoothness of $\Lambda$]\label{a:Lambda_function}
$\Lambda(\cdot)$ is twice continuously differentiable near $(\alpha_1(0,\gamma^*),\ldots,$ $\alpha_L(0,\gamma^*))$.
\end{assumption}

Assumptions~\ref{a:double_robustness}--\ref{a:Lambda_function} are the key substantive conditions. Assumption~\ref{a:double_robustness} extends double robustness from $\theta_0$ to its trimmed counterpart~$\theta_h$; we verify it for each design in Section~\ref{sec:verification}. Assumption~\ref{a:m_function}(i) is the key condition from Section~\ref{sec:dr_bc}. When the propensity score is specified correctly, this gives $\xi_\ell(0;\gamma^*)=0$. When the outcome regression model is specified correctly, the residual-containing moments satisfy $\xi_\ell(a;\gamma^*)=0$ for all $a$, so in particular $\xi_\ell(0;\gamma^*)=0$; the companion normalization moments need not satisfy this statement.
 Part~(ii) controls the polynomial approximation error. Assumption~\ref{a:Lambda_function} holds whenever the population moments $\alpha_\ell(0,\gamma^*)$ that appear as denominators in $\Lambda$ are bounded away from zero; this is a condition on the target estimand (e.g., compliance share bounded away from zero for the LATE, group share bounded away from zero for DiD), not on individual observations, and therefore remains compatible with the weak-overlap regime.

\begin{assumption}[Regularity conditions]\label{a:regularity}
The following conditions hold for each $\ell=1,\ldots,L$:
\begin{enumerate}
\item[\textnormal{(a)}] \textit{First-stage influence function.}
$\alpha_\ell(h,\hat\gamma)-\alpha_\ell(h,\gamma^*)=(\expen{\cdot}-\E[\cdot])[\phi_\ell]+o_p(n^{-1/2})$,
where $\phi_\ell$ is the influence function of $\alpha_\ell(h,\gamma)$ at $\gamma^*$.

\item[\textnormal{(b)}] \textit{Sieve estimation.}
For each $\kappa=1,\ldots,k$,
\[
\hat{\xi}_\ell^{(\kappa)}(0;\gamma^*)-{\xi}_\ell^{(\kappa)}(0;\gamma^*)-(\expen{\cdot}-\E[\cdot])[\psi_{\ell,\kappa}(\gamma^*)]=o_p(n^{-1/2}h^{1-\kappa}),
\]
where $\psi_{\ell,\kappa}(\gamma)=q_{K}^{(\kappa)}(0)'\E\bracks{q_{K}(A_\ell(\gamma))q_{K}(A_\ell(\gamma))'}^{-1}q_{K}(A_\ell(\gamma))(B_\ell(\gamma)-{\xi}_\ell(A_\ell(\gamma);\gamma))$.

\item[\textnormal{(c)}] \textit{Moment bound.}
$\E\bracks{\omega_\ell(h,\gamma^*)^2}=o(n^{1/2})$, where $\omega_\ell$ is defined in~\eqref{eq:omega_ell} below.

\item[\textnormal{(d)}] \textit{Stochastic equicontinuity.}
$\hat\alpha_\ell(h,\hat\gamma)-\alpha_\ell(h,\hat\gamma)-\hat\alpha_\ell(h,\gamma^*)+\alpha_\ell(h,\gamma^*)=o_p(n^{-1/2})$.

\item[\textnormal{(e)}] \textit{Rate condition.}
$n h^{2k}=O(1)$ as $n \rightarrow \infty$.
\end{enumerate}
\end{assumption}

\noindent Parts~(a)--(d) require the first-stage estimator, sieve regression, influence function moments, and the sample criterion to behave well; lower-level sufficient conditions (e.g., for parametric first stages) are Appendix \ref{appendix:sieve}. Part~(e) imposes an upper bound on $h$ that controls the trimming bias; a complementary lower bound of the form $n h^{4} \to \infty$, needed to control the variance of the sieve bias-correction term, is stated in Appendix~\ref{appendix:lower_bound}. 



The influence function $\omega_\ell$ of each moment $\hat\alpha_\ell$ has four components:
\begin{align}\label{eq:omega_ell}
\omega_\ell(h,\gamma)
&=\underbrace{\frac{B_\ell(\gamma)}{A_\ell(\gamma)}\idx{|A_\ell(\gamma)|\geq h}}_{\text{trimmed ratio}}
\;+\;\underbrace{\sum_{\kappa=1}^{k}\frac{A_\ell(\gamma)^{\kappa-1}\idx{|A_\ell(\gamma)|<h}}{\kappa !} \cdot {\xi}_\ell^{(\kappa)}(0;\gamma)}_{\text{bias correction}}
\nonumber\\[6pt]
&\quad+\;\underbrace{\sum_{\kappa=1}^{k}\frac{\E\bracks{A_\ell(\gamma)^{\kappa-1}\idx{|A_\ell(\gamma)|<h}}}{\kappa !} \cdot \psi_{\ell,\kappa}(\gamma)}_{\text{sieve estimation}}
\;+\;\underbrace{\phi_\ell}_{\text{first stage}}.
\end{align}
Let $\Lambda_\ell$ denote the partial derivative of $\Lambda$ with respect to its $\ell$-th argument, and define the overall influence function
\begin{equation}\label{eq:varphi}
\varphi=\sum_{\ell=1}^L\Lambda_\ell(\alpha_1(h,\gamma^*),\ldots,\alpha_L(h,\gamma^*))\,\omega_\ell(h,\gamma^*).
\end{equation}
Note that $\Lambda_\ell$ is evaluated at the population moments $\alpha_\ell(h,\gamma^*)$---the probability limits of the bias-corrected moment estimators---rather than at $\alpha_\ell(0,\gamma^*)$; these are the correct expansion points for the delta method because, under the outcome-correct route, $\alpha_\ell(h,\gamma^*)$ need not equal $\alpha_\ell(0,\gamma^*)$ for the normalization moments. Under Assumption \ref{a:double_robustness}, when $h$ goes to zero at the rate in Assumption \ref{a:regularity}(e), $\alpha_\ell(h,\gamma^*)$ converges to $\alpha_\ell(0,\gamma^*)$ and the influence function targets $\theta_0$.

\begin{theorem}\label{theorem_asy_distribution}
Under Assumptions~\ref{a:double_robustness}--\ref{a:regularity}:
\begin{enumerate}
\item[\textnormal{(i)}] $\hat\theta-\theta_0= (\expen{\cdot}-\E[\cdot])[\varphi]+o_p(n^{-1/2})$.
\item[\textnormal{(ii)}] If in addition $\E\bracks{\varphi^2}$ is bounded away from zero and ${\E\bracks{|\varphi-\E\bracks{\varphi}|^{2+\eta}}}/({n^{\eta/2}\E\bracks{(\varphi-\E\bracks{\varphi})^2}^{(2+\eta)/2}})=o(1)$ for some $\eta>0$, then
$\displaystyle(\hat\theta-\theta_0)\Big/\sqrt{\E\bracks{(\varphi-\E\bracks{\varphi})^2}/n}\convd\N(0,1).$
\end{enumerate}
\end{theorem}

\smallskip
A proof is in  Appendix \ref{appendix:proof_theorem}. A key difficulty relative to \citet{sasaki2018estimation} is that $\theta_0$ is a nonlinear function of multiple ratio moments with heterogeneous convergence rates, so the standard delta method does not directly apply. The variance is estimated by $\hat\sigma^2 = \expen{(\hat\varphi - \expen{\hat\varphi})^2}$, where $\hat\varphi$ plugs sample analogs into~\eqref{eq:omega_ell}--\eqref{eq:varphi}.

\begin{remark}[Diverging variance]\label{remark:diverging_variance}
Assumption~\ref{a:regularity}(c) allows $\E\bracks{\omega_\ell^2}$ to diverge, accommodating heavy tails from $B_\ell/A_\ell$ when $h$ is small. The Lyapunov condition ensures the CLT holds despite this.
\end{remark}

\begin{remark}[Verification across designs]
The design-specific conditions underlying Assumptions~\ref{a:double_robustness}--\ref{a:Lambda_function} are satisfied for the ATE, LATE, and staggered DiD estimands introduced in Section~\ref{sec:setup}; see Appendix Table~\ref{tab:AB_decomposition} for the corresponding $(A_\ell,B_\ell)$ decompositions and Appendix~\ref{sec:verification} for the formal verification. 
\end{remark}


{\small\singlespacing
\setlength{\bibsep}{1pt plus 0.3ex}
\putbib
}
\end{bibunit}

\newpage
\begin{bibunit}
\appendix
\newgeometry{margin=2.5cm}
\clearpage
\renewcommand\thefigure{OA-\arabic{figure}}
\renewcommand\thetable{OA-\arabic{table}}
\renewcommand*{\thepage}{OA - \arabic{page}}
\renewcommand\thesection{\Alph{section}}
\renewcommand\thesubsection{\Alph{section}.\arabic{subsection}}



\setcounter{figure}{0}
\setcounter{table}{0}
\setcounter{page}{1}

\begin{center}
	{\Large{Doubly Robust Estimators with Weak Overlap}:\\Supplemental Appendix}\\[1em]
 Yukun Ma ~~~~Pedro H.C. Sant'Anna ~~~~Yuya Sasaki~~~~Takuya Ura  \\[0.5em]

        \today \\[1em]
\end{center}








This Supplemental Appendix provides additional results for ``Doubly Robust Estimators with Weak Overlap.'' Section~\ref{appendix:simulation_dgp} presents the data-generating process for the simulation study in Section~\ref{sec:trim_problem} of the main text.
Section~\ref{sec:verification} verifies Assumptions~\ref{a:double_robustness}--\ref{a:Lambda_function} for each design from Section~\ref{sec:setup} of the main text. 
Section~\ref{appendix:did_outcomes} reports event-study estimates for additional outcomes in the Dias--Fontes DiD application discussed in Section~\ref{sec:app_did}. 
Section~\ref{appendix:ps_density} plots the propensity score distribution in Fryer-Levitt application.
Section~\ref{appendix:h_sensitivity} presents sensitivity analyses of the empirical results to the choice of trimming threshold~$h$. Sections~\ref{appendix:sieve}--\ref{appendix:propositions} contain the technical material: sieve estimation details and lower-level sufficient conditions for Assumption~\ref{a:regularity}, the proof of Theorem~\ref{theorem_asy_distribution}, and proofs of Propositions~\ref{prop:ate}--\ref{prop:did}.

\section{Simulation Design}\label{appendix:simulation_dgp}

This section provides the details of the data-generating process (DGP) used in Figure~\ref{fig:Sims_DiD}. Our design builds on \citet{Kang2007} and \citet{sant2020doubly}, adapted to a panel DiD setting with two periods and two groups.

\subsection{Data-generating process}

We generate $n = 10{,}000$ independent units. For each unit $i$, draw $X_i = (X_{i1}, X_{i2}, X_{i3}, X_{i4})'$ with each component independently Student-$t$ with 10 degrees of freedom. Define the nonlinear transformations
\[
Z_i = \bigl(Z_{i1}, Z_{i2}, Z_{i3}, Z_{i4}\bigr)' = \left(\frac{X_{i1}}{\sigma_1},\; \frac{X_{i1}^2 - X_{i2}^2}{\sigma_2},\; \frac{X_{i3}^3}{\sigma_3},\; \frac{X_{i4}^3}{\sigma_4}\right)',
\]
where $\sigma_j$ is the standard deviation of the $j$-th component (computed from the Student-$t$ moments), so that each $Z_{ij}$ has unit variance. Define two index functions:
\begin{align*}
f_{\text{reg}}(W) &= 1 + W_1 + W_2 + W_3 + W_4, \\
f_{\text{ps}}(W) &= 1.5 + W_1 + W_2 + W_3 + W_4,
\end{align*}
for a generic argument $W = (W_1, W_2, W_3, W_4)'$.

\medskip

Treatment group membership is determined by a logistic model: $D_i = \idx{p(W_i^{\text{ps}}) \geq U_i}$, where $p(w) = \exp(f_{\text{ps}}(w)) / (1 + \exp(f_{\text{ps}}(w)))$ and $U_i \sim \text{Uniform}(0,1)$. The potential outcomes follow a panel structure:
\begin{align*}
Y_{i,0} &= f_{\text{reg}}(W_i^{\text{reg}}) + \upsilon_i + \varepsilon_{i,0}, \\
Y_{i,1}(d) &= 2\,f_{\text{reg}}(W_i^{\text{reg}}) + \upsilon_i + \varepsilon_{i,1}(d), \quad d \in \{0,1\},
\end{align*}
where $\varepsilon_{i,0}$, $\varepsilon_{i,1}(0)$, $\varepsilon_{i,1}(1) \overset{\text{iid}}{\sim} \mathcal{N}(0,1)$, and $\upsilon_i \sim \mathcal{N}(D_i \cdot f_{\text{reg}}(W_i^{\text{reg}}), 1)$ captures unit-level unobserved heterogeneity that depends on the treatment status and the regression index. The observed outcomes are $Y_{i,0}$ (pre-treatment) and $Y_{i,1} = D_i\,Y_{i,1}(1) + (1-D_i)\,Y_{i,1}(0)$ (post-treatment). By construction, $\E[Y_{i,1}(1) - Y_{i,1}(0) \mid D_i = 1] = 0$, so the true ATT is zero.

\subsection{Misspecification configurations}

The key feature of this design is that $X$ and $Z$ are nonlinearly related: a researcher who uses $X$ when the true model depends on $Z$ (or vice versa) faces misspecification. We consider four configurations:

\begin{center}
\renewcommand{\arraystretch}{1.2}
\begin{tabular}{@{}llll@{}}
\toprule
Configuration & $W^{\text{ps}}$ & $W^{\text{reg}}$ & Interpretation \\
\midrule
DGP 1 & $Z$ & $Z$ & Both models correctly specified \\
DGP 2 & $X$ & $Z$ & PS misspecified, outcome correct \\
DGP 3 & $Z$ & $X$ & PS correct, outcome misspecified \\
DGP 4 & $X$ & $X$ & Both models misspecified \\
\bottomrule
\end{tabular}
\end{center}

\noindent In each configuration, the researcher observes $(X, D, Y_0, Y_1)$ and estimates the propensity score and outcome regression using $Z$ (not $X$). 

\subsection{Weak overlap}

The logistic index $f_{\text{ps}}$ with an intercept of 1.5, combined with the heavy-tailed Student-$t$ covariates, produces a non-negligible fraction of comparison-group units with propensity scores close to one. With $df = 10$ degrees of freedom, approximately 5--10\% of comparison units have $\hat{p}(X) > 0.95$, creating the weak overlap that motivates trimming.

\subsection{Estimation details}

For each Monte Carlo draw, we estimate the ATT using the DR DiD estimator of \citet{Callaway2021}:
\begin{enumerate}
\item Estimate the propensity score via logistic regression of $D$ on $Z$.
\item Estimate the outcome regression via OLS of $Y_1 - Y_0$ on $Z$ among controls ($D = 0$).
\item Compute the DR DiD ATT using the sample analogs of~\eqref{eq:dr_did}, with comparison-group weights determined by the estimated propensity score.
\end{enumerate}
We compare three estimators across a grid of trimming thresholds
\[
h \in \{0,\; 0.01,\; 0.025,\; 0.05,\; 0.10\}\text{:}
\]
(1)~the standard (untrimmed) DR estimator ($h = 0$); (2)~the trimmed DR estimator without bias correction (DR-Trim); and (3)~our proposed DR-BC estimator with $k = 1$ and $K = 3$.

Figure~\ref{fig:Sims_DiD} in the main text presents results for DGPs~2 and~3 (the two single-misspecification cases that test double robustness). Results for DGPs~1 and~4 are qualitatively similar: under DGP~1 (both models correct), all estimators perform well, and DR-BC improves precision; under DGP~4 (both models misspecified), all estimators are inconsistent, as expected. We run $10{,}000$ Monte Carlo repetitions.

\section{Design-specific verification}\label{sec:verification}

We now verify Assumptions~\ref{a:double_robustness}--\ref{a:Lambda_function} for each design from Section~\ref{sec:setup}. Table~\ref{tab:AB_decomposition} summarizes the $(A_\ell, B_\ell)$ decompositions; the resulting estimators specialize~\eqref{eq:alpha_hat} to each case.

\begin{table}[htbp]
\caption{$(A_\ell,B_\ell)$ Decompositions by Research Design}
\label{tab:AB_decomposition}
\vspace{-0.5em}
\centering
\resizebox{.55\textwidth}{!}{%
\renewcommand{\arraystretch}{1.3}%
\begin{tabular}{@{\hspace{8pt}} c @{\hphantom{mmmmmmm}} l @{\hphantom{mmmmmmmmmmmm}} l @{\hspace{8pt}}}
\toprule
$\ell$ & \multicolumn{1}{c}{$B_\ell(\gamma^*)$} & \multicolumn{1}{c}{$A_\ell(\gamma^*)$} \\
\midrule
\multicolumn{3}{c}{\textit{Panel A. Average Treatment Effect (ATE)}} \\[2pt]
1 & $D\bigl(Y-m_Y^*(1,X)\bigr)$ & $p_D^*(X)$ \\
2 & $D$ & $p_D^*(X)$ \\
3 & $(1-D)\bigl(Y-m_Y^*(0,X)\bigr)$ & $1-p_D^*(X)$ \\
4 & $1-D$ & $1-p_D^*(X)$ \\[4pt]
\midrule
\multicolumn{3}{c}{\textit{Panel B. Local Average Treatment Effect (LATE)}} \\[2pt]
1 & $Z\bigl(Y-m_Y^{*\text{LATE}}(1,X)\bigr)$ & $p_Z^*(X)$ \\
2 & $Z$ & $p_Z^*(X)$ \\
3 & $(1-Z)\bigl(Y-m_Y^{*\text{LATE}}(0,X)\bigr)$ & $1-p_Z^*(X)$ \\
4 & $1-Z$ & $1-p_Z^*(X)$ \\
5 & $Z\bigl(D-m_D^{*\text{LATE}}(1,X)\bigr)$ & $p_Z^*(X)$ \\
6 & $Z$ & $p_Z^*(X)$ \\
7 & $(1-Z)\bigl(D-m_D^{*\text{LATE}}(0,X)\bigr)$ & $1-p_Z^*(X)$ \\
8 & $1-Z$ & $1-p_Z^*(X)$ \\[4pt]
\midrule
\multicolumn{3}{c}{\textit{Panel C. Difference-in-Differences (DiD)}} \\[2pt]
1 & $\idx{G=g}\bigl(Y_t-Y_{g-\delta-1}-m_{g,t}^*(X)\bigr)$ & $1$ \\
2 & $p_{g,t}^*(X)\,C_{g,t}\bigl(Y_t-Y_{g-\delta-1}-m_{g,t}^*(X)\bigr)$ & $1-p_{g,t}^*(X)$ \\
3 & $\idx{G=g}$ & $1$ \\
4 & $p_{g,t}^*(X)\,C_{g,t}$ & $1-p_{g,t}^*(X)$ \\[2pt]
\bottomrule
\end{tabular}}
\end{table}

\begin{example}[ATE under unconfoundedness]\label{example:ate}
The ATE decomposition uses $L = 4$; see Table~\ref{tab:AB_decomposition}. Weak overlap arises when $p_D(X)$ is near zero or one.

\begin{prop}\label{prop:ate}
Suppose: \textnormal{(i)}~for each $d\in\{0,1\}$,
\[
\E\bracks{p_D(X)^{d}\,(1-p_D(X))^{1-d}\bigl(m_Y(d,X)-m_Y^*(d,X)\bigr) \;\big|\; p_D^*(X)=1-d}=0;
\]
and \textnormal{(ii)}~the functions
\begin{align*}
s&\mapsto \E\bracks{p_D(X)\bigl(m_Y(1,X)-m_Y^*(1,X)\bigr) \mid p_D^*(X)=s}, \\
s&\mapsto \E\bracks{(1-p_D(X))\bigl(m_Y(0,X)-m_Y^*(0,X)\bigr) \mid 1-p_D^*(X)=s}
\end{align*}
are $(k+1)$-times continuously differentiable near~$0$. Then Assumptions~\ref{a:double_robustness}--\ref{a:Lambda_function} hold for the ATE.
\end{prop}

The bias-corrected ATE estimator is
$\hat\theta = \expen{\hat m_Y(1,X) - \hat m_Y(0,X)} + \hat\alpha_1/\hat\alpha_2 - \hat\alpha_3/\hat\alpha_4$,
where each $\hat\alpha_\ell = \hat\alpha_\ell(h,\hat\gamma)$ is as in~\eqref{eq:alpha_hat}. Trimming discards observations with $\hat p_D(X) < h$ or $1 - \hat p_D(X) < h$.
\end{example}

\begin{example}[LATE under instrumental variables]\label{example:late}
The LATE decomposition uses $L = 8$, with moments $\ell = 1,\ldots,4$ for the reduced form and $\ell = 5,\ldots,8$ for the first stage; see Table~\ref{tab:AB_decomposition}. The function $\Lambda$ is the Wald ratio.

\begin{prop}\label{prop:late}
Suppose: \textnormal{(i)}~for each $z\in\{0,1\}$ and each $R \in \{Y,D\}$,
\[
\E\bracks{p_Z(X)^{z}\,(1-p_Z(X))^{1-z}\bigl(m_R^{\text{LATE}}(z,X)-m_R^{*\text{LATE}}(z,X)\bigr) \;\big|\; p_Z^*(X)=1-z}=0;
\]
\textnormal{(ii)}~the functions
\begin{align*}
s&\mapsto \E\bracks{p_Z(X)\bigl(m_R^{\text{LATE}}(1,X)-m_R^{*\text{LATE}}(1,X)\bigr) \mid p_Z^*(X)=s}, \\
s&\mapsto \E\bracks{(1-p_Z(X))\bigl(m_R^{\text{LATE}}(0,X)-m_R^{*\text{LATE}}(0,X)\bigr) \mid 1-p_Z^*(X)=s}
\end{align*}
are $(k+1)$-times continuously differentiable near~$0$ for each $R \in \{Y,D\}$; and \textnormal{(iii)}~the compliance share $\E\bracks{D(1)-D(0)}$ is bounded away from zero. Then Assumptions~\ref{a:double_robustness}--\ref{a:Lambda_function} hold for the LATE.
\end{prop}

The structure parallels the ATE case, with the instrument $Z$ replacing $D$ and the instrument propensity score $p_Z(X)$ replacing $p_D(X)$. Condition~(i) holds under either DR condition: when $p_Z^* = p_Z$, the boundary $p_Z(X) = 0$ or $p_Z(X) = 1$ forces $Z$ or $1-Z$ to zero a.s., so $B_\ell = 0$; when $m_R^{*\text{LATE}} = m_R^{\text{LATE}}$, the outcome residual in $B_\ell$ has conditional mean zero given $X$, so $\xi_\ell(a;\gamma^*) = 0$ for all $a$. Condition~(iii) ensures the Wald ratio denominator is well-defined.

The bias-corrected LATE estimator is the Wald ratio of two bias-corrected DR expressions:
\[
\hat\theta = \frac{\expen{\hat m_Y^{\text{LATE}}(1,X)-\hat m_Y^{\text{LATE}}(0,X)}+\hat\alpha_1/\hat\alpha_2-\hat\alpha_3/\hat\alpha_4}{\expen{\hat m_D^{\text{LATE}}(1,X)-\hat m_D^{\text{LATE}}(0,X)}+\hat\alpha_5/\hat\alpha_6-\hat\alpha_7/\hat\alpha_8}.
\]
The ITT effect is the numerator alone.
\end{example}

\begin{example}[Staggered DiD]\label{example:did}
The DiD ATT$(g,t)$ decomposition uses $L = 4$, with $A_1 = A_3 = 1$ (trivial denominators); only $\alpha_2$ and $\alpha_4$ involve weak overlap through $A_\ell = 1 - p_{g,t}(X)$. See Table~\ref{tab:AB_decomposition}. Staggered DiD identification relies on irreversible treatment, $\delta$-limited anticipation, and conditional parallel trends as discussed in \citet{Callaway2021}. Let $\delta \geq 0$ denote the number of anticipation periods. The key assumptions are: (i)~irreversibility ($D_{t-1} = 1 \Rightarrow D_t = 1$ a.s.); (ii)~limited anticipation (for $t < g - \delta$, $Y_t(g) = Y_t(\infty)$ a.s.); and (iii)~conditional parallel trends (for the relevant comparison group $\mathcal{C}_{g,t}$, $\E\bracks{Y_t(\infty)-Y_{g-\delta-1}(\infty) \mid X, G=g} = \E\bracks{Y_t(\infty)-Y_{g-\delta-1}(\infty) \mid X, C_{g,t}=1}$ a.s.). One can rely on never-treated or not-yet-treated comparison groups, and on weighted versions of these assumptions as discussed in \citet{baker2025difference}; we omit the details to avoid repetition.

\begin{prop}\label{prop:did}
Suppose that $0< \E\bracks{\idx{G=g}} < 1$ and: \textnormal{(i)}
\[
\E\bracks{C_{g,t}\bigl(m_{g,t}(X)-m_{g,t}^*(X)\bigr) \mid p^*_{g,t}(X)=1}=0;
\]
 and \textnormal{(ii)}~the relevant conditional expectations are $(k+1)$-times continuously differentiable near~$0$. Then Assumptions~\ref{a:double_robustness}--\ref{a:Lambda_function} hold for $\text{ATT}(g,t)$ and for $\text{ES}(e) = \sum_g w_{g,e}^{\text{es}} \cdot \text{ATT}(g,g+e)$.
\end{prop}

When $p_{g,t}^* = p_{g,t}$, $p_{g,t}(X) = 1$ implies $C_{g,t} = 0$ a.s., so $B_2 = B_4 = 0$. When $m_{g,t}^* = m_{g,t}$, the residual moment $\ell=2$ has conditional mean zero among comparison units, so $\xi_2(0;\gamma^*) = 0$. The companion normalization moment $\ell=4$ need not satisfy this stronger statement.

The bias-corrected estimator for each group-time ATT is
\begin{align*}
\hat\theta_{g,t} = \frac{\expen{\idx{G=g}(Y_t - Y_{g-\delta-1} - \hat m_{g,t}(X))}}{\expen{\idx{G=g}}} - \frac{\hat\alpha_2(h,\hat\gamma_{g,t})}{\hat\alpha_4(h,\hat\gamma_{g,t})},
\end{align*}
where $\hat\alpha_2$ and $\hat\alpha_4$ trim on $1 - \hat p_{g,t}(X) < h$. Event-study parameters aggregate across groups: $\hat\theta(e) = \sum_g \hat w_{g,e} \hat\theta_{g,g+e}$, with standard errors from $\hat\varphi(e) = \sum_g \hat w_{g,e} \hat\varphi_{g,g+e}$, where inference is conducted conditional on the observed cohort shares (i.e., treating $\hat w_{g,e}$ as fixed). Simultaneous confidence bands are constructed via multiplier bootstrap following \citet{Callaway2021}. 
\end{example}
\section{Additional DiD Outcomes: Dias--Fontes Application}\label{appendix:did_outcomes}

\begin{figure}[htbp]
\centering
\includegraphics[width=0.75\textwidth]{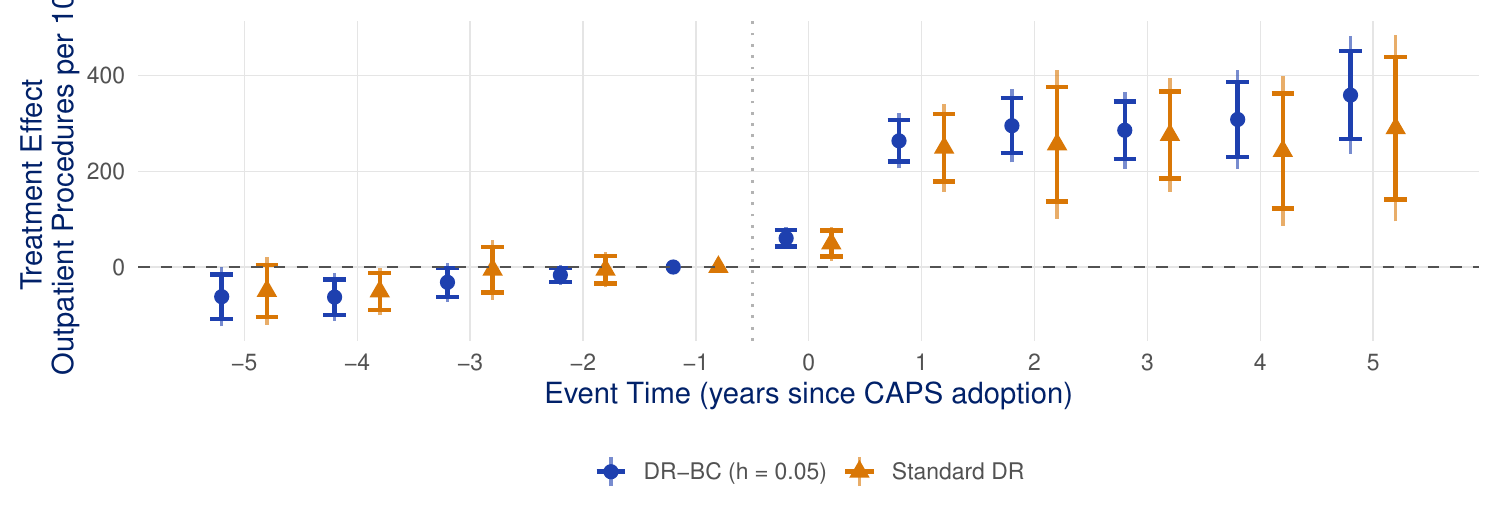}
\caption{Effect of CAPS adoption on outpatient mental health procedures per 10,000 population}
\label{fig:DiD_pa_total}
\justifying
\vspace{0.1cm}\noindent\scriptsize{\textit{Notes.} Event-study estimates using the DR DiD framework of \citet{Callaway2021} with not-yet-treated municipalities as the comparison group and no anticipation ($\delta=0$). DR-BC is our proposed estimator with $h=0.05$; Standard DR is the untrimmed estimator. Points denote point estimates, vertical bars denote 95\% simultaneous confidence bands, and error bars with caps denote 95\% pointwise confidence intervals. Data from \citet{DiasFontes2024_AEJPolicy}. Sample starts in 2008.}
\end{figure}

\begin{figure}[htbp]
\centering
\includegraphics[width=0.75\textwidth]{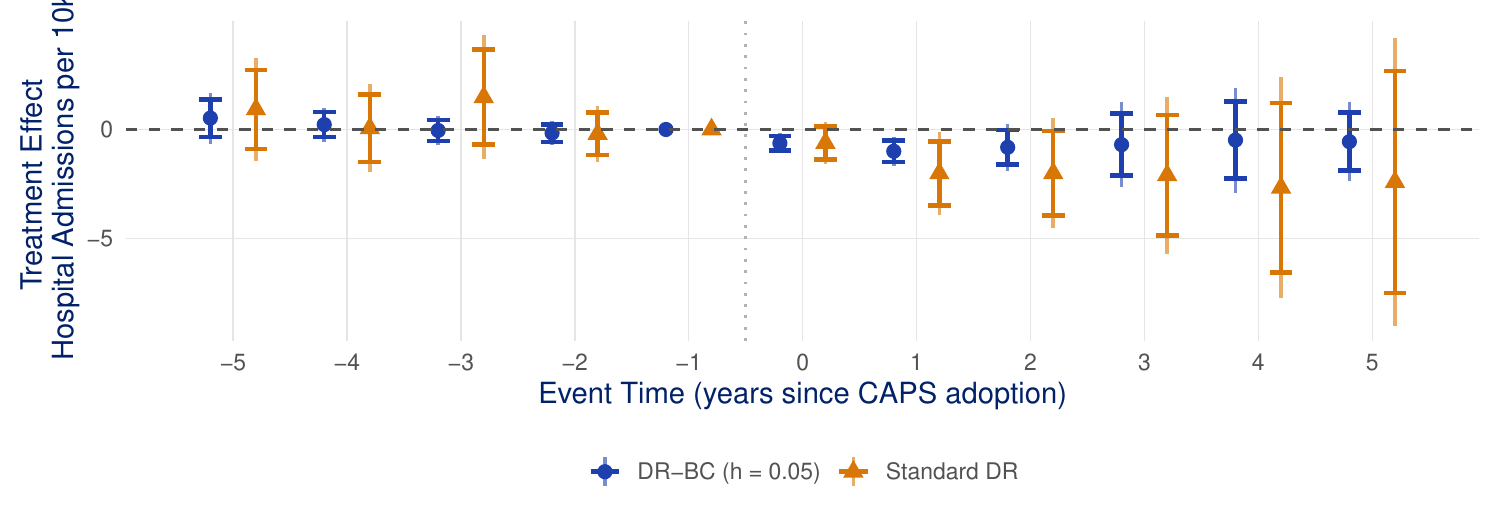}
\caption{Effect of CAPS adoption on mental health hospital admissions per 10,000 population}
\label{fig:DiD_sih_tnet_F}
\justifying
\vspace{0.1cm}\noindent\scriptsize{\textit{Notes.} Event-study estimates using the DR DiD framework of \citet{Callaway2021} with not-yet-treated municipalities as the comparison group and no anticipation ($\delta=0$). DR-BC is our proposed estimator with $h=0.05$; Standard DR is the untrimmed estimator. Points denote point estimates, vertical bars denote 95\% simultaneous confidence bands, and error bars with caps denote 95\% pointwise confidence intervals. Data from \citet{DiasFontes2024_AEJPolicy}. Sample starts in 2002.}
\end{figure}

\begin{figure}[htbp]
\centering
\includegraphics[width=0.75\textwidth]{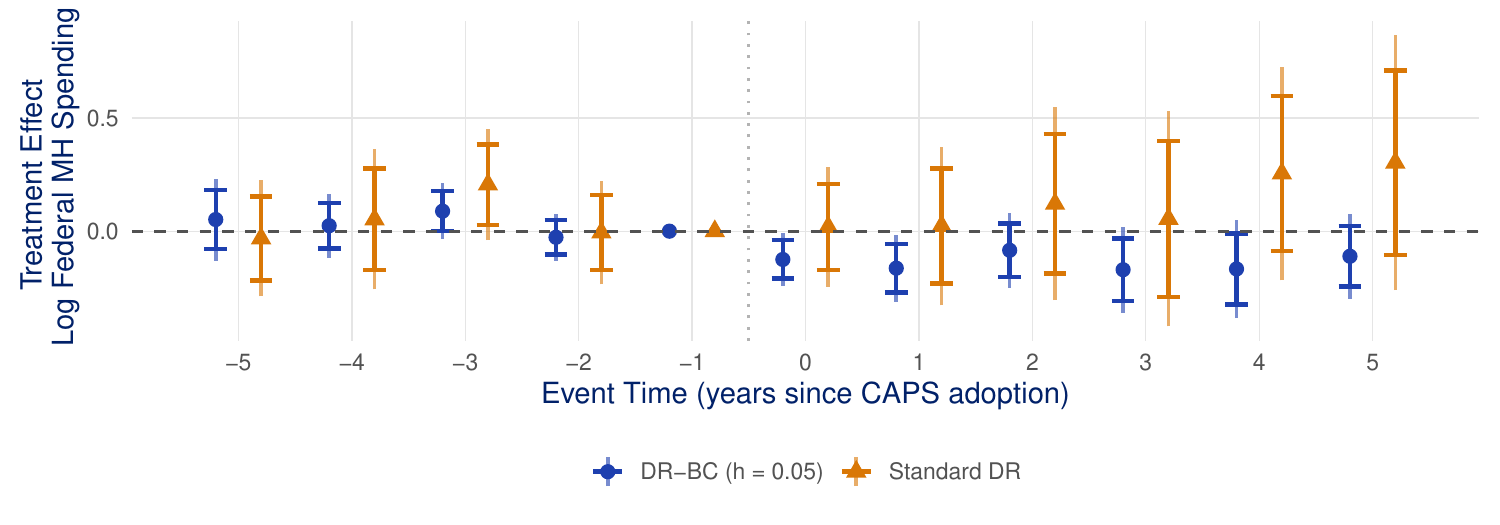}
\caption{Effect of CAPS adoption on log federal mental health hospital spending}
\label{fig:DiD_lnvalortotal}
\justifying
\vspace{0.1cm}\noindent\scriptsize{\textit{Notes.} Event-study estimates using the DR DiD framework of \citet{Callaway2021} with not-yet-treated municipalities as the comparison group and no anticipation ($\delta=0$). DR-BC is our proposed estimator with $h=0.05$; Standard DR is the untrimmed estimator. Points denote point estimates, vertical bars denote 95\% simultaneous confidence bands, and error bars with caps denote 95\% pointwise confidence intervals. Data from \citet{DiasFontes2024_AEJPolicy}. Sample starts in 2002.}
\end{figure}

\begin{figure}[htbp]
\centering
\includegraphics[width=0.75\textwidth]{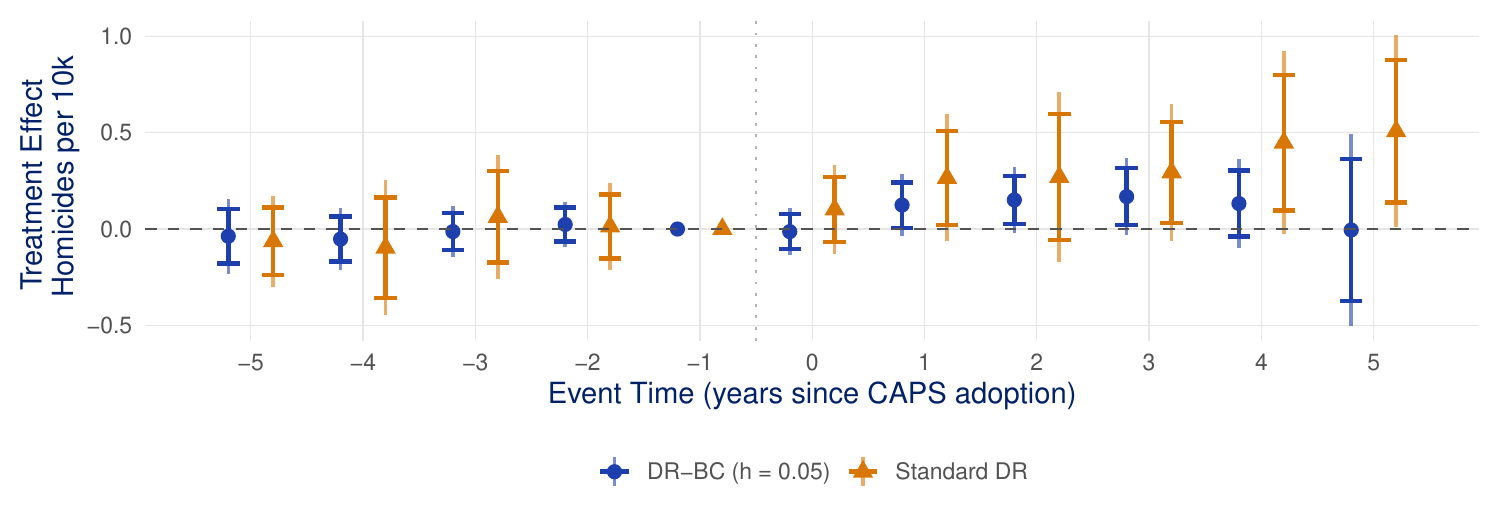}
\caption{Effect of CAPS adoption on homicides per 10,000 population}
\label{fig:DiD_sim_agressao}
\justifying
\vspace{0.1cm}\noindent\scriptsize{\textit{Notes.} Event-study estimates using the DR DiD framework of \citet{Callaway2021} with not-yet-treated municipalities as the comparison group and no anticipation ($\delta=0$). DR-BC is our proposed estimator with $h=0.05$; Standard DR is the untrimmed estimator. Points denote point estimates, vertical bars denote 95\% simultaneous confidence bands, and error bars with caps denote 95\% pointwise confidence intervals. Data from \citet{DiasFontes2024_AEJPolicy}. Sample starts in 2002.}
\end{figure}

The main paper presents event-study estimates for mental health practitioners (Figure~\ref{fig:DiD_applications}, Panel~(b)), where we set $\delta=1$ because the original analysis of \citet{DiasFontes2024_AEJPolicy} finds evidence of anticipation for that outcome. For the remaining four outcomes, figures \ref{fig:DiD_pa_total}--\ref{fig:DiD_sim_agressao}, no such evidence is present, so we set $\delta=0$. In all specifications, the comparison group consists of not-yet-treated municipalities, and the covariate set includes state fixed effects and 29 baseline characteristics.

\section{Propensity Score Distribution: Fryer--Levitt Application}\label{appendix:ps_density}

\begin{figure}[htbp]
\centering
\includegraphics[width=0.75\textwidth]{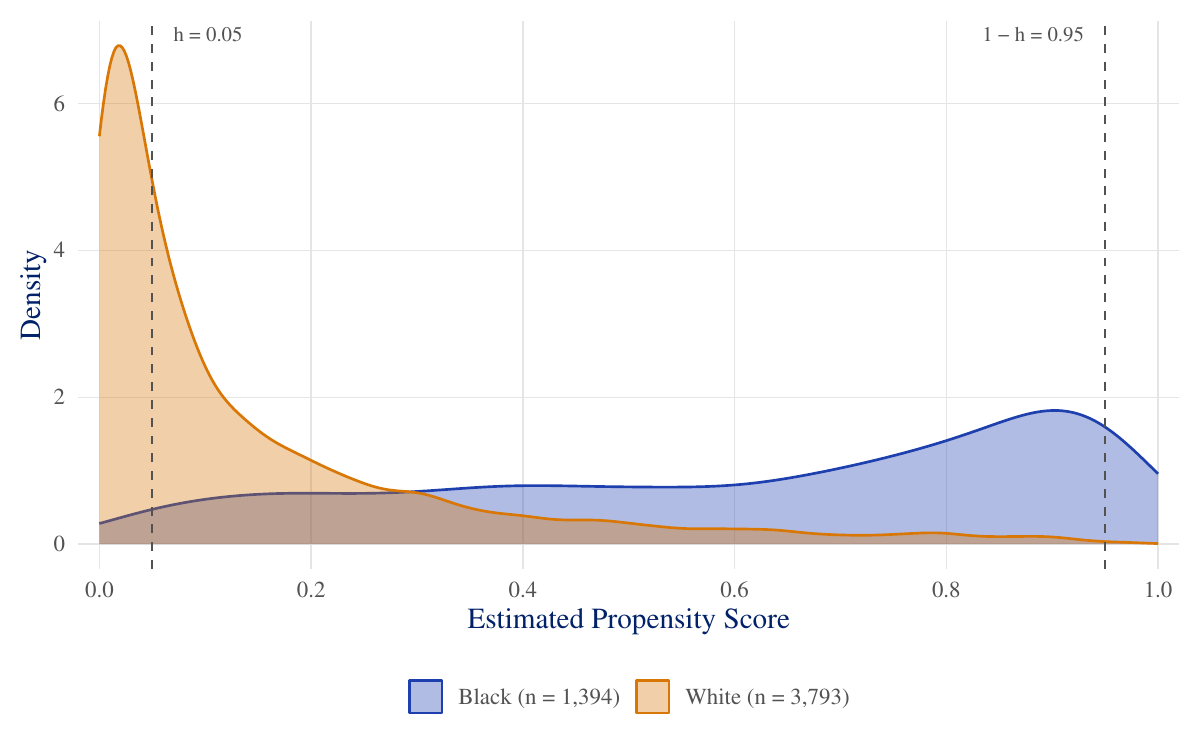}
\caption{Propensity score distribution: Black--White test score gap (Fryer--Levitt)}
\label{fig:ps_density_fryer}
\justifying
\vspace{0.1cm}\noindent\scriptsize{\textit{Notes.} Density of estimated propensity scores from a logit model of race (Black vs.\ White) on the full set of covariates used by \citet{FryerLevitt2013_AER}, including interviewer fixed effects. Dashed vertical lines mark the trimming thresholds $h = 0.05$ and $1-h = 0.95$. At $h = 0.05$, 36 observations (0.69\%) at 9 months and 38 (0.73\%) at 24 months are trimmed — the ``active arm'' of the trimmed moment: treated observations with $\hat p(X) < 0.05$ or control observations with $\hat p(X) > 0.95$.}
\end{figure}

Figure~\ref{fig:ps_density_fryer} shows the estimated propensity score distribution for the Fryer--Levitt application. The interviewer fixed effects create many small cells in which only one race is observed, pushing the propensity score toward its boundaries for a non-negligible share of observations. The trimming thresholds $h = 0.05$ and $1 - h = 0.95$ are indicated by dashed vertical lines.

\section{Sensitivity to \texorpdfstring{$h$}{h}}\label{appendix:h_sensitivity}

\begin{figure}[htbp]
\caption{Sensitivity to $h$: Black--White Test Score Gap (Fryer--Levitt)}
\label{fig:h_sensitivity_ATE}
\begin{center}
\includegraphics[width=\textwidth]{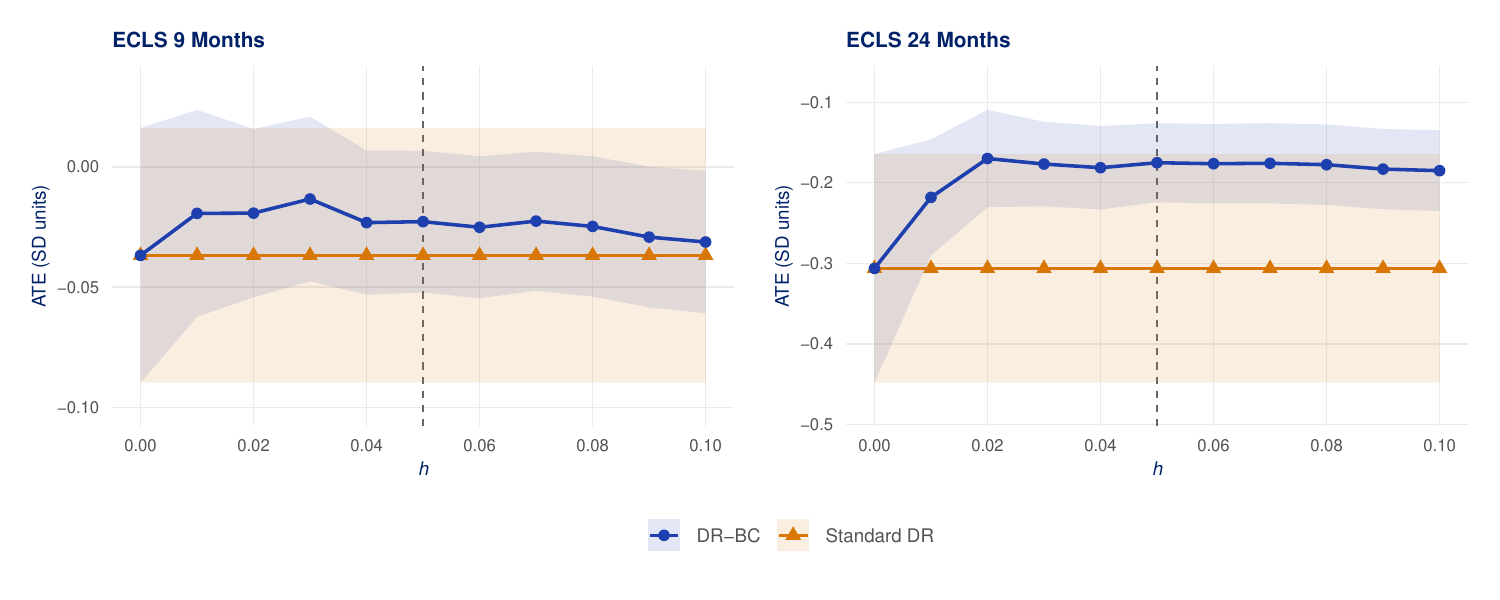}
\end{center}
\justifying
\vspace{0.1cm}\noindent\scriptsize{\textit{Notes.} DR-BC (blue) and Standard DR (gold) ATE estimates with 95\% confidence intervals (shaded bands) as a function of $h\in[0,0.10]$. The dashed vertical line marks $h=0.05$, the recommended default. DR-BC estimates are approximately flat for $h\geq 0.03$, while the Standard DR estimate (at $h=0$) has a much wider confidence interval. At 24 months, the SE drops from 0.072 to 0.025 as $h$ increases from 0 to 0.05.}
\end{figure}

\begin{figure}[htbp]
\caption{Sensitivity to $h$: 401(k) Eligibility on Net Financial Assets}
\label{fig:h_sensitivity_LATE}
\begin{center}
\includegraphics[width=\textwidth]{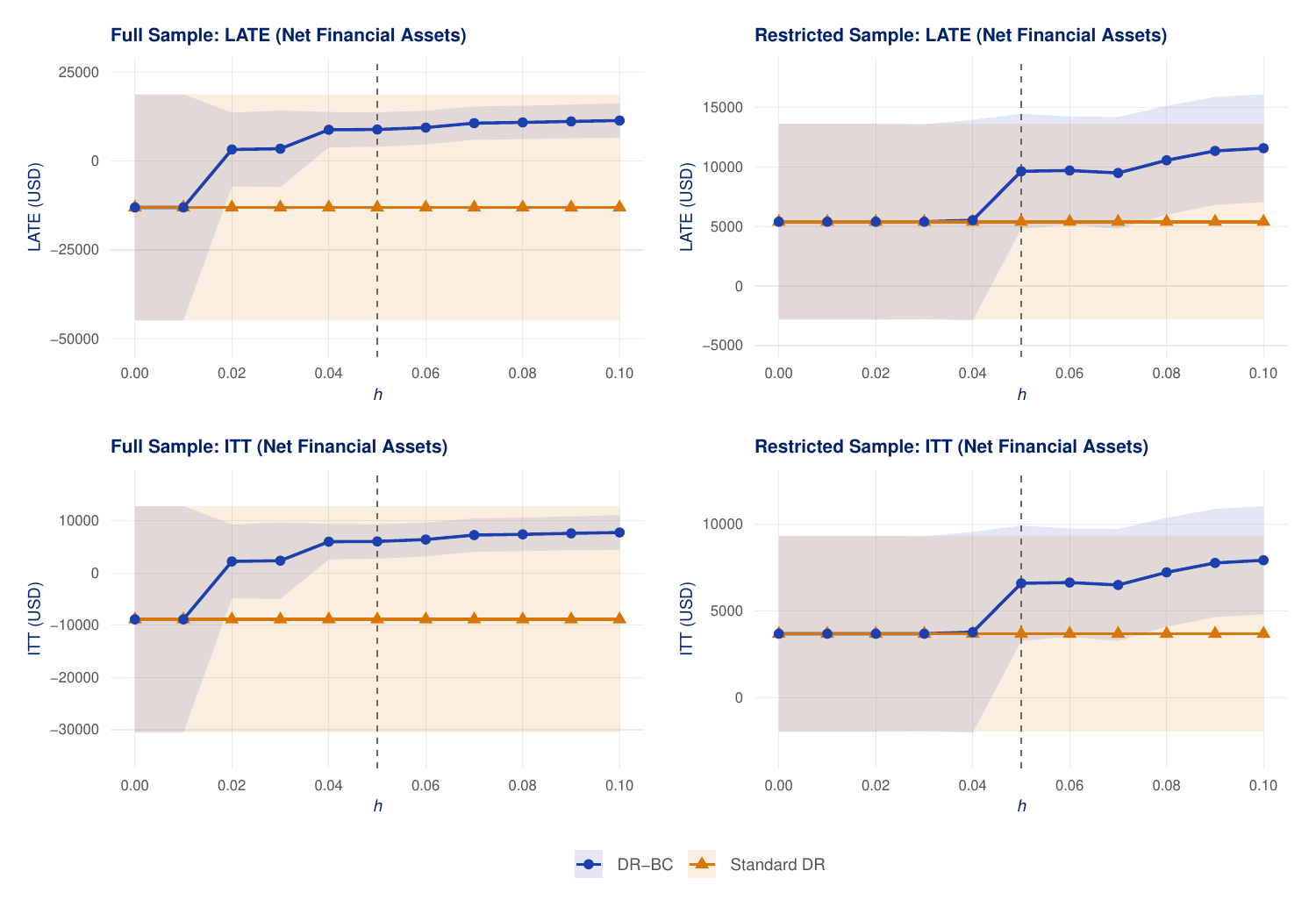}
\end{center}
\justifying
\vspace{0.1cm}\noindent\scriptsize{\textit{Notes.} DR-BC (blue) and Standard DR (gold) estimates with 95\% confidence intervals as a function of $h\in[0,0.10]$ for the 401(k) application. Top row: LATE; bottom row: ITT. Left column: full sample ($n=9{,}910$, positive income); right column: restricted sample ($n=9{,}275$, income in [\$10k, \$200k]). The dashed vertical line marks $h=0.05$. The Standard DR estimate is constant across $h$ (it does not use trimming). DR-BC estimates stabilize for $h\geq 0.03$, with precision gains largest in the full sample where overlap is weakest.}
\end{figure}

\begin{figure}[htbp]
\caption{Sensitivity to $h$: Medicaid Expansion on Mortality (Baker et al.)}
\label{fig:h_sensitivity_Medicaid}
\begin{center}
\includegraphics[width=0.55\textwidth]{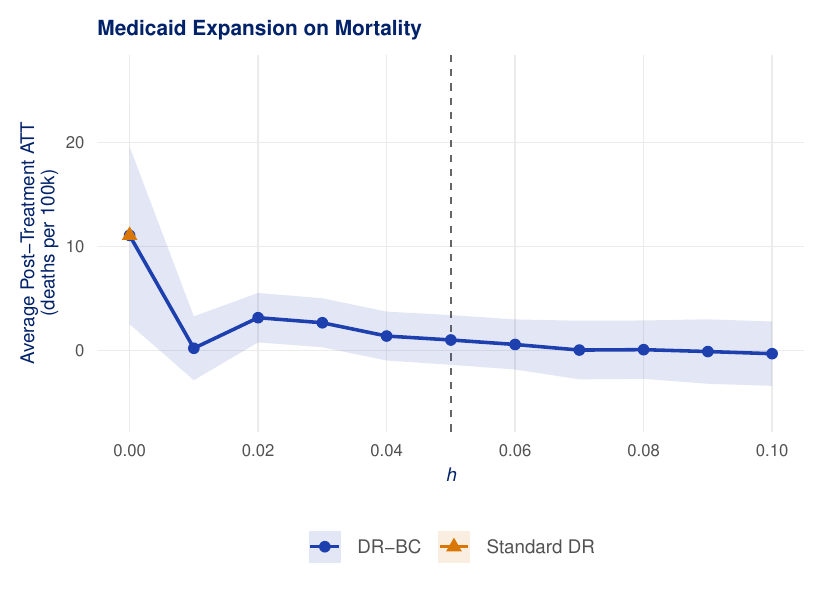}
\end{center}
\justifying
\vspace{0.1cm}\noindent\scriptsize{\textit{Notes.} Average post-treatment ATT for the crude mortality rate (ages 20--64, deaths per 100,000), computed as the mean of DR-BC event-study estimates across post-treatment event times. Never-treated counties serve as the comparison group, with population weights. The dashed vertical line marks $h=0.05$. The Standard DR estimate (gold) is constant across $h$. DR-BC estimates are approximately stable across $h$, with tighter confidence intervals than Standard DR.}
\end{figure}

\begin{figure}[htbp]
\caption{Sensitivity to $h$: Dias--Fontes DiD (Mental Health Practitioners, $\delta=1$)}
\label{fig:h_sensitivity_DiD}
\begin{center}
\includegraphics[width=0.55\textwidth]{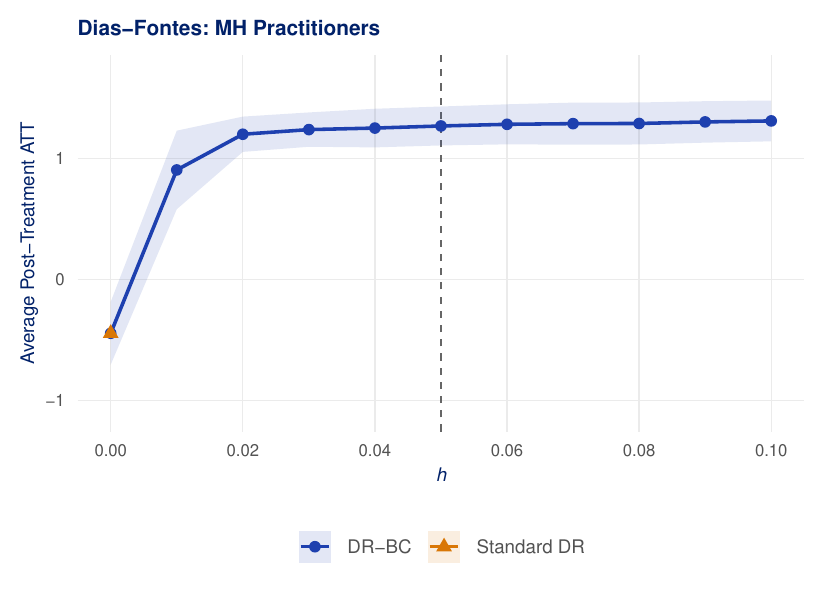}
\end{center}
\justifying
\vspace{0.1cm}\noindent\scriptsize{\textit{Notes.} Average post-treatment ATT for mental health practitioners per 10,000 population with 1-year anticipation ($\delta=1$), computed as the mean of DR-BC event-study estimates across post-treatment event times. The dashed vertical line marks $h=0.05$. The Standard DR estimate (gold) is constant across $h$. DR-BC estimates are approximately stable for $h\geq 0.03$.}
\end{figure}

Figures~\ref{fig:h_sensitivity_ATE}--\ref{fig:h_sensitivity_DiD} display estimates and 95\% confidence intervals as functions of $h\in[0, 0.10]$ for each application.

A common pattern across applications is a substantial change in point estimates between $h = 0$ (the standard untrimmed DR estimator) and small positive $h$ (e.g., $h = 0.02$ or $h = 0.03$), followed by approximate stability for $h \geq 0.03$. This pattern is expected and informative: the jump reflects the removal of a small number of observations with extreme weights that dominate the untrimmed estimator, while the subsequent stability indicates that the bias correction is adequately reconstructing the trimmed observations' contribution across a range of thresholds. It is important to distinguish two notions of ``trimming.'' The share of observations whose arm-specific IPW weight is unstable---treated units with $\hat{p}(X) < h$ or control units with $\hat{p}(X) > 1-h$---is what actually drives the variance of the standard DR estimator and the sharp initial jump. This share is small across applications: 36--38 observations (0.69--0.73\%) in Fryer--Levitt, 2 ineligible observations with effective weights above 30 in the full-sample 401(k), and 177 comparison observations (0.8\% on average, up to 3.3\% in the most affected group-time cell) in Medicaid. The share of observations whose covariate cell places them in the trimmed region is typically larger---as in Fryer--Levitt, where 35--38\% of the sample shares a covariate cell with a high-weight unit---because trimming operates at the covariate-cell level. These additional observations carry near-unit effective weights and would not destabilize the standard DR estimator on their own; they are reconstructed by the bias correction to preserve the target estimand. The relevant diagnostic is not whether the DR-BC estimate at $h = 0.05$ matches the standard DR estimate at $h = 0$---it generally will not, since the standard DR estimate is inflated by extreme weights---but whether the DR-BC estimate is stable across $h$ for moderate values of $h$. Stability across $h \in [0.03, 0.10]$ suggests the polynomial approximation is adequate and the results are not driven by the specific choice of threshold. Instability across this range would suggest model misspecification or that the conditions underlying Assumption~\ref{a:m_function} are not met, and researchers should exercise caution in such cases.

\section{Sieve Estimation Details}\label{appendix:sieve}

The sieve estimator for $\xi_\ell(\cdot;\gamma)$ uses the shifted orthonormal Legendre polynomial basis of degree $K$:
\begin{align*}
q_K(a)=\bigl(1,\;\sqrt{3}(2a-1),\;\sqrt{5}(6a^2-6a+1),\;\sqrt{7}(20a^3-30a^2+12a-1),\;\ldots\bigr)'.
\end{align*}
These polynomials are orthonormal on $[0,1]$, making $q_K$ natural for propensity scores. The sieve estimator for the $\kappa$-th derivative at zero is
\begin{align*}
\hat{\xi}_\ell^{(\kappa)}(0;\gamma)
=q_K^{(\kappa)}(0)'\,\expen{q_K(A_\ell(\gamma))q_K(A_\ell(\gamma))'}^{-1}\expen{q_K(A_\ell(\gamma))B_\ell(\gamma)}.
\end{align*}
The influence function of this sieve estimator, centered at its population analog $\xi_\ell^{(\kappa)}(0;\gamma^*)$, is
\begin{align*}
\psi_{\ell,\kappa}(\gamma^*)
=q_K^{(\kappa)}(0)'\,\E\bracks{q_K(A_\ell(\gamma^*))q_K(A_\ell(\gamma^*))'}^{-1}q_K(A_\ell(\gamma^*))\bigl(B_\ell(\gamma^*)-\xi_\ell(A_\ell(\gamma^*);\gamma^*)\bigr),
\end{align*}
which is the $\psi_{\ell,\kappa}(\gamma^*)$ appearing in Assumption~\ref{a:regularity}(b) and the influence function~\eqref{eq:omega_ell}.

\medskip

The following lemma, based on \citet{belloni2015some}, gives lower-level sufficient conditions for Assumption~\ref{a:regularity}(b). See also \citet{chen2015optimal}.

\begin{lemma}\label{lem:sieve}
For each $\ell=1,\ldots,L$ and $\kappa=1,\ldots,k$, suppose:
\begin{enumerate}
\item[\textnormal{(i)}] The eigenvalues of $\E\bracks{q_K(A_\ell(\gamma^*))q_K(A_\ell(\gamma^*))'}$ are bounded above and away from zero.
\item[\textnormal{(ii)}] $\sqrt{\log K}\,(K + K^{5/2-s})\,\|q_K^{(\kappa)}(0)\| = o(h^{1-\kappa}n^{1/2})$, where $s$ is the smoothness order of $\xi_\ell(\cdot;\gamma^*)$.
\item[\textnormal{(iii)}] $K^{1-s}\|q_K^{(\kappa)}(0)\| = o(h^{1-\kappa})$.
\item[\textnormal{(iv)}] The sieve approximation error $r_{K,\ell}^{(\kappa)}(0) = o(h^{1-\kappa}n^{-1/2})$, where
\begin{align*}
r_{K,\ell}^{(\kappa)}(0)
= \xi_\ell^{(\kappa)}(0;\gamma^*) - q_K^{(\kappa)}(0)'\E\bracks{q_K(A_\ell(\gamma^*))q_K(A_\ell(\gamma^*))'}^{-1}\E\bracks{q_K(A_\ell(\gamma^*))\xi_\ell(A_\ell(\gamma^*);\gamma^*)}.
\end{align*}
\end{enumerate}
Then Assumption~\ref{a:regularity}\textnormal{(b)} holds for the index $(\ell,\kappa)$. 
\end{lemma}

\medskip

\noindent For parametric first-stage estimators $\hat\gamma$ (e.g., logistic propensity score, linear outcome regression), Assumption~\ref{a:regularity}(a) and (d) hold under standard regularity conditions \citep[see, e.g.,][]{newey1997convergence}. Specifically, if $\hat\gamma$ is root-$n$ consistent with influence function $\phi$, and if $\alpha_\ell(h,\gamma)$ is differentiable in $\gamma$ with bounded derivative, then (a) holds with $\phi_\ell = (\partial/\partial\gamma')\alpha_\ell(h,\gamma^*)\cdot\phi$, and (d) follows from a uniform law of large numbers.

\section{Lower Bound on the Trimming Rate}\label{appendix:lower_bound}

Assumption~\ref{a:regularity}(c) requires $\E\bracks{\omega_\ell(h,\gamma^*)^2} = o(n^{1/2})$. The following lemma gives a lower-level sufficient condition that yields the complementary rate condition $nh^4\to\infty$.

\begin{lemma}\label{lem:omega_bound}
Let $\ell$ be any index with $1\leq\ell\leq L$. Suppose $\E\bracks{B_\ell(\gamma^*)^2}$, $\xi_\ell^{(\kappa)}(0;\gamma^*)$ for each $\kappa=1,\ldots,k$, and $\E\bracks{\|\phi\|^2}$ are all bounded. If $nh^4\to\infty$, $\|(\partial/\partial\gamma')\alpha_\ell(h,\gamma^*)\|=o(n^{1/4})$, and $\E\bracks{\psi_{\ell,\kappa}(\gamma^*)^2}=o(n^{1/2})$, then Assumption~\ref{a:regularity}\textnormal{(c)} holds.
\end{lemma}

\begin{proof}
By the triangle inequality applied to the four-component decomposition~\eqref{eq:omega_ell}:
\begin{align*}
\E\bracks{\omega_\ell(h,\gamma^*)^2}^{1/2}
&\leq h^{-1}\E\bracks{B_\ell(\gamma^*)^2}^{1/2}
+\sum_{\kappa=1}^{k}\frac{h^{\kappa-1}}{\kappa!}\,|\xi_\ell^{(\kappa)}(0;\gamma^*)|
\\&\quad
+\sum_{\kappa=1}^{k}\frac{h^{\kappa-1}}{\kappa!}\,\E\bracks{\psi_{\ell,\kappa}(\gamma^*)^2}^{1/2}
+\Bigl\|\frac{\partial}{\partial\gamma'}\alpha_\ell(h,\gamma^*)\Bigr\|\E\bracks{\|\phi\|^2}^{1/2}.
\end{align*}
The first term is $O(h^{-1})$. Since $nh^4\to\infty$ implies $h\to 0$ at most as fast as $n^{-1/4}$, we have $h^{-1}=o(n^{1/4})$. The remaining terms are $o(n^{1/4})$ by assumption. Hence $\E\bracks{\omega_\ell(h,\gamma^*)^2}^{1/2}=o(n^{1/4})$, which gives $\E\bracks{\omega_\ell(h,\gamma^*)^2}=o(n^{1/2})$.
\end{proof}

\noindent The condition $nh^4\to\infty$ is a lower bound on $h$. Writing the bounds with their constants, the feasible range for $k=1$ is $c_1\, n^{-1/4} < h < c_2\, n^{-1/2}$, where $c_1$ depends on the moment bounds of $\omega_\ell$ and $c_2$ depends on the smoothness of $\xi_\ell$. Since $n^{-1/4} \gg n^{-1/2}$ for large $n$, this range is asymptotically empty. However, for finite $n$ the range may be non-empty if $c_2/c_1$ is sufficiently large: at $n = 10{,}000$, the range becomes $0.10\,c_1 < h < 0.01\,c_2$, which is non-empty whenever $c_2 > 10\,c_1$. The constants are DGP-dependent and not directly estimable with current methods, so a data-driven shrinking-$h$ rule remains an open problem. For $k \geq 3$, the upper bound $h = O(n^{-1/(2k)})$ is less restrictive (e.g., $h = O(n^{-1/6})$ for $k=3$), and the feasible range $c_1\, n^{-1/4} < h < c_2\, n^{-1/(2k)}$ is non-degenerate. In principle, this permits a vanishing-$h$ analysis with $k \geq 3$, though in our simulations higher-order bias corrections degrade finite-sample performance because the required sieve derivatives $\xi_\ell^{(\kappa)}(0)$ for $\kappa \geq 3$ are imprecisely estimated with a fixed low-degree basis (Section~\ref{sec:dr_bc}).

\section{Proof of Theorem~\ref{theorem_asy_distribution}}\label{appendix:proof_theorem}

We establish the four intermediate results:
\begin{align}
&\alpha_\ell(h,\gamma^*)-\alpha_\ell(0,\gamma^*)=o(n^{-1/2}),\label{ap:step1}\\
&\hat\alpha_\ell(h,\hat\gamma)-\alpha_\ell(h,\gamma^*)=(\expen{\cdot}-\E[\cdot])[\omega_\ell(h,\gamma^*)]+o_p(n^{-1/2}),\label{ap:step2}\\
&\hat\alpha_\ell(h,\hat\gamma)-\alpha_\ell(0,\gamma^*)=o_p(n^{-1/4}),\label{ap:step3}\\
&\hat\theta-\theta_0=(\expen{\cdot}-\E[\cdot])[\varphi]+o_p(n^{-1/2}).\label{ap:step4}
\end{align}
Statement (i) of the Theorem~\ref{theorem_asy_distribution} is~\eqref{ap:step4}. Statement (ii) follows immediately: by the Lyapunov CLT applied to $(\expen{\cdot}-\E[\cdot])[\varphi]$ and the fact that $\E\bracks{\varphi^2}$ is bounded away from zero, combining with~\eqref{ap:step4} gives $(\hat\theta-\theta_0)/\sqrt{\E\bracks{(\varphi-\E\bracks{\varphi})^2}/n}\convd\N(0,1)$.

\medskip
\noindent\textbf{Step 1: Proof of~\eqref{ap:step1}.}

By the definition of $\alpha_\ell$, the law of iterated expectations, and a $k$th-order Taylor expansion of $\xi_\ell(a;\gamma^*)$ around $a=0$:
\begin{align*}
&\alpha_\ell(h,\gamma^*)-\alpha_\ell(0,\gamma^*)
\\
&= -\E\bracks{\frac{B_\ell(\gamma^*)}{A_\ell(\gamma^*)}\idx{|A_\ell(\gamma^*)|<h}}
+\sum_{\kappa=1}^{k}\frac{\E\bracks{A_\ell(\gamma^*)^{\kappa-1}\idx{|A_\ell(\gamma^*)|<h}}}{\kappa!}\xi_\ell^{(\kappa)}(0;\gamma^*)
\\
&= -\E\bracks{\frac{\xi_\ell(A_\ell(\gamma^*);\gamma^*)}{A_\ell(\gamma^*)}\idx{|A_\ell(\gamma^*)|<h}}
+\sum_{\kappa=1}^{k}\frac{\E\bracks{A_\ell(\gamma^*)^{\kappa-1}\idx{|A_\ell(\gamma^*)|<h}}}{\kappa!}\xi_\ell^{(\kappa)}(0;\gamma^*)
\\
&= -\frac{\E\bracks{A_\ell(\gamma^*)^{k}\int_0^1(1-u)^{k}\xi_\ell^{(k+1)}(uA_\ell(\gamma^*);\gamma^*)\,du\cdot\idx{|A_\ell(\gamma^*)|<h}}}{k!},
\end{align*}
where the second equality uses the law of iterated expectations $\E[\cdot]=\E[\E[\cdot\mid A_\ell(\gamma^*)]]$, and the third uses Assumption~\ref{a:m_function}. For residual moments (odd-indexed $\ell$ in the ATE and LATE designs, or the outcome-residual moments in DiD), $\xi_\ell(0;\gamma^*)=0$ under either DR route, so the Taylor expansion around $a=0$ eliminates all terms through order $k$ and $\alpha_\ell(h,\gamma^*)-\alpha_\ell(0,\gamma^*)=o(h^k)$. For normalization moments (even-indexed $\ell$ in ATE and LATE, or the comparison-group share in DiD), $\xi_\ell(0;\gamma^*)=0$  holds under the propensity-score route, yielding the same $o(h^k)$ bound. Under the outcome regression route ($m^*=m$, $p^*\neq p$), however, $\xi_\ell(0;\gamma^*)\neq 0$, so $\alpha_\ell(h,\gamma^*)-\alpha_\ell(0,\gamma^*)$ need not vanish. This does not compromise the estimand because the corresponding residual moments are identically zero along this route, making $\Lambda$ insensitive to the normalization moments. It does, however, require the delta-method expansion in Step~4 to be centered at $\alpha(h, \gamma^*)$ rather than $\alpha(0,\gamma^*)$.
Since  $\xi_\ell$ is $(k+1)$-times differentiable near zero, the $k$th-order Taylor remainder with integral form is
\begin{align*}
\xi_\ell(a;\gamma^*)=\sum_{\kappa=1}^{k}\frac{a^\kappa}{\kappa!}\xi_\ell^{(\kappa)}(0;\gamma^*)+\frac{a^{k+1}}{k!}\int_0^1(1-u)^{k}\xi_\ell^{(k+1)}(ua;\gamma^*)\,du.
\end{align*}
Since $|A_\ell(\gamma^*)|\leq h$ on the indicator event and $\xi_\ell^{(k+1)}$ is bounded near zero,
\begin{align*}
\alpha_\ell(h,\gamma^*)-\alpha_\ell(0,\gamma^*)=O\!\left(h^k\,\E\bracks{\idx{|A_\ell(\gamma^*)|<h}}\right)=o(n^{-1/2}),
\end{align*}
where the last step uses Assumption~\ref{a:regularity}(e): $nh^{2k}=O(1)$ implies $h^k=O(n^{-1/2})$, and the probability of the trimmed region goes to zero.

\medskip
\noindent\textbf{Step 2: Proof of~\eqref{ap:step2}.}

By Assumption~\ref{a:regularity}(d),
\begin{align*}
\hat\alpha_\ell(h,\hat\gamma)-\alpha_\ell(h,\gamma^*)
=\hat\alpha_\ell(h,\gamma^*)-\alpha_\ell(h,\gamma^*)
+\alpha_\ell(h,\hat\gamma)-\alpha_\ell(h,\gamma^*)
+o_p(n^{-1/2}).
\end{align*}
By Assumptions~\ref{a:regularity}(a) and (b):
\begin{align*}
\hat\alpha_\ell(h,\hat\gamma)-\alpha_\ell(h,\gamma^*)
&= \hat\alpha_\ell(h,\gamma^*)-\alpha_\ell(h,\gamma^*)
+\frac{\partial}{\partial\gamma'}\alpha_\ell(h,\gamma^*)\cdot(\expen{\cdot}-\E[\cdot])[\phi]
+o_p(n^{-1/2})\\
&= (\expen{\cdot}-\E[\cdot])[\omega_\ell(h,\gamma^*)]+o_p(n^{-1/2}),
\end{align*}
where the last equality collects the three stochastic terms from the trimmed ratio, bias correction, and sieve estimation parts of $\hat\alpha_\ell$ into $\omega_\ell(h,\gamma^*)$ as defined in~\eqref{eq:omega_ell}.

\medskip
\noindent\textbf{Step 3: Proof of~\eqref{ap:step3}.}

Combining~\eqref{ap:step1} and~\eqref{ap:step2}:
\begin{align*}
\hat\alpha_\ell(h,\hat\gamma)-\alpha_\ell(0,\gamma^*)
=(\expen{\cdot}-\E[\cdot])[\omega_\ell(h,\gamma^*)]+o_p(n^{-1/2}).
\end{align*}
By Assumption~\ref{a:regularity}(c), $\E\bracks{\omega_\ell(h,\gamma^*)^2}=o(n^{1/2})$, so $(\expen{\cdot}-\E[\cdot])[\omega_\ell(h,\gamma^*)]=o_p(n^{-1/4})$ by Markov's inequality. Hence $\hat\alpha_\ell(h,\hat\gamma)-\alpha_\ell(0,\gamma^*)=o_p(n^{-1/4})$.

\medskip
\noindent\textbf{Step 4: Proof of~\eqref{ap:step4}.}

By a first-order Taylor expansion of $\Lambda$ around $(\alpha_1(h,\gamma^*),\ldots,\alpha_L(h,\gamma^*))$ under Assumption~\ref{a:Lambda_function}:
\begin{align*}
\hat\theta-\theta_h(\gamma^*)
&= \Lambda(\hat\alpha_1(h,\hat\gamma),\ldots,\hat\alpha_L(h,\hat\gamma))
  -\Lambda(\alpha_1(h,\gamma^*),\ldots,\alpha_L(h,\gamma^*))
\\
&= \sum_{\ell=1}^L\Lambda_\ell(\alpha_1(h,\gamma^*),\ldots,\alpha_L(h,\gamma^*))\bigl(\hat\alpha_\ell(h,\hat\gamma)-\alpha_\ell(h,\gamma^*)\bigr)
  +O_p\!\Bigl(\sum_{\ell=1}^L|\hat\alpha_\ell(h,\hat\gamma)-\alpha_\ell(h,\gamma^*)|^2\Bigr).
\end{align*}
By~\eqref{ap:step3}, each $|\hat\alpha_\ell(h,\hat\gamma)-\alpha_\ell(h,\gamma^*)|^2=o_p(n^{-1/2})$.  
By Assumption \ref{a:double_robustness}, $\theta_h(\gamma^*)=\theta_0+o(h^k)$ and $nh^{2k}=O(1)$ gives the trimming bias as $o(n^{-1/2})$. Substituting~\eqref{ap:step2} into the linear term yields
\begin{align*}
\hat\theta-\theta_0
&=\sum_{\ell=1}^L\Lambda_\ell(\alpha_1(h,\gamma^*),\ldots,\alpha_L(h,\gamma^*))\cdot(\expen{\cdot}-\E[\cdot])[\omega_\ell(h,\gamma^*)]
+o_p(n^{-1/2})\\
&=(\expen{\cdot}-\E[\cdot])[\varphi]+o_p(n^{-1/2}),
\end{align*}
where the last equality uses the definition of $\varphi$ in~\eqref{eq:varphi}. This completes the proof.~\qed

\section{Proofs of Propositions~\ref{prop:ate}--\ref{prop:did}}\label{appendix:propositions}

Each proposition simultaneously verifies Assumptions~\ref{a:double_robustness}--\ref{a:Lambda_function}. This merges two steps that in earlier work (for a single-design estimator) appeared separately: (i) showing $\xi_\ell(0;\gamma^*)=0$, which gives Assumption~\ref{a:m_function}(i); and (ii) showing $\theta_h(\gamma^*)=\theta_0+o(h^k)$ under either DR condition, which is Assumption~\ref{a:double_robustness}. We establish both here for each design.

\medskip
\noindent\textbf{Proof of Proposition~\ref{prop:ate}.}

\textbf{Part 1: Verifying Assumption~\ref{a:m_function}.}

For $\ell=1$ ($A_1=p_D^*(X)$, $B_1=D(Y-m_Y^*(1,X))$), the law of iterated expectations gives
\begin{align*}
\xi_1(s;\gamma^*)
&= \E\bracks{D\bigl(Y-m_Y^*(1,X)\bigr)\mid p_D^*(X)=s}.
\end{align*}
Part~(i) of the proposition requires $\xi_1(0;\gamma^*)=0$ under either DR condition. When $p_D^*=p_D$: the boundary $p_D^*(X)=0$ forces $D=0$ a.s., so $B_1=0$ and $\xi_1(0;\gamma^*)=0$. When $m_Y^*=m_Y$: the misspecification residual $m_Y(1,X)-m_Y^*(1,X)=0$, so $\xi_1(0;\gamma^*)=s\cdot 0=0$. Condition~(ii) implies $s\mapsto\E\bracks{p_D^*(X)(m_Y(1,X)-m_Y^*(1,X))\mid p_D(X)=s}$ is $(k+1)$-times continuously differentiable near zero, which gives Assumption~\ref{a:m_function}(ii) for $\ell=1$. For $\ell=2$ ($B_2=D$, $A_2=p_D^*(X)$), $\xi_2(0;\gamma^*)=\E\bracks{D\mid p_D^*(X)=0}=0$ when $p_D^*=p_D$ (since $D=0$ a.s.\ at the boundary). When $m_Y^*=m_Y$, no separate verification for $\ell=2$ is needed because the companion numerator moment $\ell=1$ is already identically zero. The argument for $\ell=3,4$ is symmetric with $1-p_D^*(X)$ replacing $p_D^*(X)$.

\textbf{Part 2: Verifying Assumption~\ref{a:double_robustness}.}

From Step~1 of the proof of Theorem~\ref{theorem_asy_distribution} (specifically~\eqref{ap:step1}), we have for each $\ell$:
\begin{align*}
\alpha_\ell(h,\gamma^*)-\alpha_\ell(0,\gamma^*)
= -\frac{\E\bracks{A_\ell(\gamma^*)^{k}\int_0^1(1-u)^{k}\xi_\ell^{(k+1)}(uA_\ell(\gamma^*);\gamma^*)\,du\cdot\idx{|A_\ell(\gamma^*)|<h}}}{k!}.
\end{align*}
For the ATE with $\ell=2$: when $p_D^*=p_D$, $\xi_2(s;\gamma^*)=\E\bracks{D \mid p_D(X)=s}=s$, which is exactly linear, so $\xi_2^{(\kappa)}(\cdot;\gamma^*)=0$ for all $\kappa\geq 2$; the Taylor remainder therefore vanishes for any $k\geq 1$, and the expression above equals zero. When $m_Y^*=m_Y$, no separate Taylor argument is needed for $\ell=2$, because the companion numerator moment $\ell=1$ is already identically zero. For $\ell=1$: when $p_D^*=p_D$, $\xi_1(s;\gamma^*)=s\,\E\bracks{m_Y(1,X)-m_Y^*(1,X)\mid p_D(X)=s}$ has a bounded $(k+1)$-th derivative, and the expression is $O(h^k\Pr(p_D(X)<h))=o(h^k)$; when $m_Y^*=m_Y$, $\xi_1(\cdot;\gamma^*)=0$ identically, so the remainder vanishes. 
Similarly for $\ell=3,4$. Under $p_D^*=p_D$: since $\Lambda(a_1,a_2,a_3,a_4)=\E[m_Y^*(1,X)-m_Y^*(0,X)]+a_1/a_2-a_3/a_4$ is differentiable (Part~3) and each $\alpha_\ell(h,\gamma^*)-\alpha_\ell(0,\gamma^*)=o(h^k)$, It follows that $\theta_h(\gamma^*)=\theta_0+o(h^k)$. Under $m_Y^*=m_Y$: the residual moments $\ell=1,3$ are identically zero for all $h$, so the ATE representation is unchanged by trimming and $\theta_h(\gamma^*)=\theta_0$ exactly. Each route gives Assumption \ref{a:double_robustness}.

\textbf{Part 3: Verifying Assumption~\ref{a:Lambda_function}.}
$\Lambda(a_1,a_2,a_3,a_4)=\E[m_Y^*(1,X)-m_Y^*(0,X)]+a_1/a_2-a_3/a_4$ is infinitely differentiable when the denominators $a_2=\alpha_2(0,\gamma^*)$ and $a_4=\alpha_4(0,\gamma^*)$ are bounded away from zero. Under either DR condition, these denominators are strictly positive: when $p_D^*=p_D$, $\alpha_2(0,\gamma^*)=\E\bracks{D/p_D(X)}=1$; when $m_Y^*=m_Y$, $\alpha_2(0,\gamma^*)=\E\bracks{p_D(X)/p_D^*(X)}>0$ since both propensity scores are strictly between 0 and 1.~\qed

\medskip
\noindent\textbf{Proof of Proposition~\ref{prop:late}.}

The LATE decomposition uses $L=8$ moments: $\ell\in\{1,2,3,4\}$ for the reduced-form Wald numerator and $\ell\in\{5,6,7,8\}$ for the first-stage Wald denominator. In both blocks, $A_\ell\in\{p_Z^*(X),1-p_Z^*(X)\}$, mirroring the ATE structure with $Z$ replacing $D$ and $p_Z^*(X)$ replacing $p_D^*(X)$.

\textbf{Part 1: Verifying Assumption~\ref{a:m_function}.}

For odd-indexed moments ($\ell=1,3$ in the reduced form, $\ell=5,7$ in the first stage), $B_\ell$ contains an outcome or treatment residual. Taking $\ell=1$ ($B_1=Z(Y-m_Y^{*\text{LATE}}(1,X))$, $A_1=p_Z^*(X)$) as representative:
\begin{align*}
\xi_1(s;\gamma^*)
= \E\bracks{Z(m_Y^{\text{LATE}}(1,X)-m_Y^{*\text{LATE}}(1,X))\mid p_Z^*(X)=s}.
\end{align*}
When $p_Z^*=p_Z$: the boundary $p_Z^*(X)=0$ forces $Z=0$ a.s., so $B_1=0$ and $\xi_1(0;\gamma^*)=0$. When $m_Y^{*\text{LATE}}=m_Y^{\text{LATE}}$: the misspecification residual is zero, so $\xi_1(\cdot;\gamma^*)=0$ identically. The argument for $\ell=3,5,7$ is analogous (with $1-p_Z^*(X)$ replacing $p_Z^*(X)$ for $\ell=3,7$, and $D$ replacing $Y$ for $\ell=5,7$). Condition~(ii) of the proposition gives $(k+1)$-fold differentiability for the odd-indexed moments.

For even-indexed moments ($\ell=2,4,6,8$), the identity  $\xi_\ell(0;\gamma^*)=0$ is invoked only along the correctly specified propensity score path. For example, if $\ell=2$,  ($B_2=Z$, $A_2=p_Z^*(X)$): $\xi_2(s;\gamma^*)=\E\bracks{Z\mid p_Z^*(X)=s}=s$, which is exactly linear, so $\xi_2(0;\gamma^*)=0$ and $\xi_2^{(\kappa)}(0;\gamma^*)=0$ for all $\kappa\geq 2$. Under the alternative double-robustness route, $m_R^{*\text{LATE}}=m_R^{\text{LATE}}$, the odd-indexed residual moments satisfy $\xi_\ell(a;\gamma^*)=0$ for all $a$, so $\alpha_\ell(h,\gamma^*)=0$ regardless of $h$. The even-indexed normalization moments need not equal zero when the propensity score is misspecified. The numerator moments of the Wald ratio are zero regardless of $h$, so the bias-correction terms vanish irrespective of the value of  $\alpha_{\text{even}}(h,\gamma^*)$. So no boundary expansion for the even-indexed moment is needed for the Wald-ratio limit. The argument for $\ell=4,6,8$ is symmetric. 

\textbf{Part 2: Verifying Assumption~\ref{a:double_robustness}.}
When $p_Z^*=p_Z$, the Taylor argument from Step~1 applies to both the reduced-form and first-stage moments, including their companion denominator moments, so $\theta_h^{\mathrm{num}}(\gamma^*)=\theta_0^{\mathrm{num}}+o(h^k)$ and $\theta_h^{\mathrm{den}}(\gamma^*)=\theta_0^{\mathrm{den}}+o(h^k)$. The Wald ratio $\theta_h(\gamma^*)=\theta_0+o(h^k)$. When  $m_Y^{*\text{LATE}}=m_Y^{\text{LATE}}$, the odd-indexed residual moments are identically zero. Hence no separate boundary expansion is needed for the even-indexed denominator moments, and the Wald ratio is unchanged by trimming. Therefore, the Wald ratio $\theta_h(\gamma^*)=\theta_0$.

\textbf{Part 3: Verifying Assumption~\ref{a:Lambda_function}.}
$\Lambda = (\text{numerator})/(\text{denominator})$ is infinitely differentiable when the first-stage effect $\E\bracks{D(1)-D(0)}>0$, which is condition~(iii).~\qed

\medskip
\noindent\textbf{Proof of Proposition~\ref{prop:did}.}

For the DiD ATT$(g,t)$, only $\ell=2$ and $\ell=4$ involve potentially weak overlap through $A_2=A_4=1-p_{g,t}^*(X)$; $\ell=1,3$ have $A_\ell=1$, so $0\notin \supp {A_\ell}$. These moments are unaffected by trimming.

\textbf{Part 1: Verifying Assumption~\ref{a:m_function} for $\ell=2$.}

With $B_2=p_{g,t}^*(X)C_{g,t}(Y_t-Y_{g-\delta-1}-m_{g,t}^*(X))$ and $A_2=1-p_{g,t}^*(X)$, the law of iterated expectations gives
\begin{align*}
\xi_2(s;\gamma^*)
&= \E\bracks{p_{g,t}^*(X)C_{g,t}\bigl(Y_t-Y_{g-\delta-1}-m_{g,t}^*(X)\bigr)\mid 1-p_{g,t}^*(X)=s}.
\end{align*}
Since $p_{g,t}^*(X)=1-s$ is fixed on the conditioning event, this equals
\begin{align*}
\xi_2(s;\gamma^*) = (1-s)\,\E\bracks{C_{g,t}\bigl(Y_t-Y_{g-\delta-1}-m_{g,t}^*(X)\bigr)\mid 1-p_{g,t}^*(X)=s}.
\end{align*}
Condition~(i) requires $\xi_2(0;\gamma^*)=0$. When $p_{g,t}^*=p_{g,t}$: the boundary $p_{g,t}^*(X)=1$ implies $C_{g,t}=0$ a.s.\ (no comparison units at $p=1$), so $B_2=0$ and $\xi_2(0;\gamma^*)=0$. When $m_{g,t}^*=m_{g,t}$: the law of iterated expectations gives
\begin{align*}
\E\bracks{C_{g,t}\bigl(Y_t-Y_{g-\delta-1}-m_{g,t}(X)\bigr)\mid 1-p_{g,t}^*(X)=s}=0,
\end{align*}
since $\E\bracks{Y_t-Y_{g-\delta-1}-m_{g,t}(X)\mid X,C_{g,t}=1}=0$ by definition of $m_{g,t}$.  Condition~(ii) gives $(k+1)$-fold differentiability. The argument for $\ell=4$ is analogous when $p_{g,t}^*=p_{g,t}$; when $m_{g,t}^*=m_{g,t}$, no separate argument for $\ell=4$ is needed because the companion residual moment $\alpha_2(h,\gamma^*)$ is already zero for every $h$.

\textbf{Part 2: Verifying Assumption~\ref{a:double_robustness}.}

Under the correct propensity score route, the Taylor remainder argument of Step~1 applies to $\alpha_2$ and $\alpha_4$ separately, giving $\alpha_\ell(h,\gamma^*)-\alpha_\ell(0,\gamma^*)=o(h^k)$ for $\ell\in\{2,4\}$.
Under the correct outcome regression route, $\alpha_2(h,\gamma^*)=0$ for all $h$, so the ratio $\alpha_2/\alpha_4=0$ regardless of the value of $\alpha_4(h,\gamma^*)$, and Assumption~\ref{a:double_robustness} holds without requiring a Taylor argument for $\alpha_4$.

\textbf{Part 3: Verifying Assumption~\ref{a:Lambda_function}.}
$\Lambda$ for ATT$(g,t)$ is $\alpha_1/\alpha_3-\alpha_2/\alpha_4$, differentiable when $\alpha_3=\E\bracks{\idx{G=g}}>0$ (from $0<\E\bracks{\idx{G=g}}<1$) and $\alpha_4>0$.

For the event-study aggregation $\text{ES}(e)=\sum_g w_{g,e}^{\text{es}}\cdot\text{ATT}(g,g+e)$, the result follows by a weighted-sum argument over each component ATT$(g,t)$ for which the above holds.~\qed


{\small\singlespacing
\setlength{\bibsep}{1pt plus 0.3ex}
\putbib
}
\end{bibunit}
\end{document}